**Dynamic plasmonic color generation enabled by functional materials**

Frank Neubrech,[1,3] Xiaoyang Duan,[3,1] Na Liu[2,3,4*]

[1]Max Planck Institute for Intelligent Systems, Heisenbergstrasse 3, 70569 Stuttgart, Germany.

[2]Max Planck Institute for Solid State Research, Heisenbergstrasse 1, 70569 Stuttgart, Germany.

[3]Kirchhoff-Institute for Physics, University of Heidelberg, Im Neuenheimer Feld 227, 69120 Heidelberg, Germany.

[4]Centre for Advanced Materials, University of Heidelberg, Im Neuenheimer Feld 225, 69120 Heidelberg, Germany.

[*]Corresponding author. E-mail: na.liu@kip.uni-heidelberg.de

**Abstract**

**Displays are an indispensable medium to visually convey information in our daily life. Although conventional dye-based color displays have been rigorously advanced by world leading companies, critical issues still remain. For instance, color-fading and wavelength-limited resolution restrict further developments. Plasmonic colors emerging from resonant interactions between light and metallic nanostructures can overcome such restrictions. With dynamic characteristics enabled by functional materials, dynamic plasmonic coloration may find a variety of applications in display technologies. In this review, we elucidate basic concepts for dynamic plasmonic color generation and highlight recent advances. In particular, we devote our review to a selection of dynamic controls endowed by functional materials, including magnesium, liquid crystals, electrochromic polymers, phase change materials, among others. Also, we discuss their performance in view of potential applications in current display technologies.**

# 1. Introduction

Color displays are ubiquitous in our daily life. They are integral components of modern devices such as mobile phones, television, e-readers, electronic billboards, notebooks, tablets, video walls and many others, without which today's world would be hardly imaginable. The majority of the pixelated color displays integrated in these devices are emissive and backlit displays. Colors of backlit displays are generated by filtering particular wavelengths from a spectrally broad backlight utilizing dyes or pigments (passive coloration). Such devices typically involve light emitting diode (LED) backlit illumination and liquid crystal display (LCD) technology. In contrast, colors from emissive displays are directly produced by a narrowband emission from one or more light sources (active coloration), such as LEDs or organic LEDs (OLEDs). Emissive displays, in particular OLED displays, can be very thin and flexible. They exhibit excellent contrasts and consume low power (*1*). The latter is of utmost importance for mobile devices, as the displays account for the majority of power consumption.

Reduced power consumptions, wide viewing angles, and high visibility under daylight conditions make reflective displays especially attractive, in particular, for outdoor use (*2*). Different from emissive and backlit technology, reflective displays including reflective LCDs and various electrochromic screens, utilize ambient light for coloration. This special type of passive coloration offers significant advantages. First, under bright light (*e.g.* sunlight), which severely limits the visibility performance of emissive and backlit displays, their reflective counterparts offer excellent readability. Second, light emission that is power consuming is not required, since only the wavelengths of the reflected light need to be controlled. While reflective black/white electrophoretic displays (*3*), *e.g.* e-papers, have already been commercialized, the development of full-color reflective displays, including different kinds of electrochromic surfaces, still face substantial challenges (*4*). For instance, electrochromic materials integrated in reflective color displays possess excellent electrical control over their optical properties and the visual appearance (*5*). However, a low absolute reflectivity and technical challenges associated with the implementation of sub-pixels give rise to dull colors and hamper the applicability for commercial products. The situation becomes particularly challenging, if high resolution displays with subwavelength pixel sizes are desired. For pixel sizes on the micrometerscale and below, the dyes' absorption/emission dramatically decreases due to the reduced dye volumes. This holds true for reflective-type as well as for backlit-type and emissive-type displays. Additionally, the incorporated dyes bleach upon continuous UV exposure and the colors fade over time (*6*).

Structural colors, well known from coloration in nature (*7*), can overcome these limitations. Different from dyes and pigments, structural colors are generated by the interaction of light with micro- and nanostructures. Vibrant colors can be produced with the same materials (*e.g.* metals or dielectrics) by changing the geometries, dimensions or arrangements of the structures through the fabrication process or even post fabrication (*8*). Compared to pigment or dye-based coloration, colors created in this case are much brighter due to their inherently high scattering/absorption efficiencies. As a result, thin layers, or more precisely tiny volumes, are sufficient for brilliant coloration. The benefit of such small coloration volumes is obvious. Ultra-high resolution images composed of subwavelength pixels with sizes down to the smallest coloring unit, *e.g.* a single micro- or nanostructure, can be printed (*9*). Additionally, structural colors do not fade over time but provide a basically everlasting coloration due to the stability of the coloring structures. These appealing advantages have attracted great interest and stimulated intensive research on various structural coloration schemes based on metal nanostructures, dielectric metasurfaces, photonic crystals, and Fabry-Perot (FP) resonances (*6*, *10–17*).

Metal nanostructures offer the unique possibility to tailor the transmitted and reflected light by plasmons. Plasmons are collective oscillations of free electrons in metals (*18*). They provide strongly enhanced electromagnetic near-fields associated with sustainably enhanced absorption and scattering at distinct resonant wavelengths. The resonance wavelengths and thus the associated plasmonic color are determined by the material, the geometric properties of the nanostructures, their surrounding media and their arrangement, if applicable. All parameters, are nearly freely adjustable, if the appropriate fabrication approach, *e.g.* bottom-up or top-down is chosen. This freedom has led to a large number of studies on plasmonic color generation and printing. Various types of plasmons, *e.g.* surface plasmon polaritons (SPPs) or localized surface plasmon resonances (LSPRs), and related concepts, *e.g.* plasmon hybridization or coupling between plasmons and other optical excitations, have been employed to generate vibrant non-fading colors with high brightness and contrasts (*6*, *10–12*).

In this review, we highlight the recent advances in the field of dynamic plasmonic color generation and discuss their performance in view of potential applications in current display technologies. We restrict ourselves to reversible systems, in which the dynamic control of the plasmonic colors is enabled by a functional material. We first outline key performance indicators, which we use later to discuss the strengths and weaknesses of the respective coloration approaches. We briefly introduce the basic physical processes for adjusting the plasmonic resonances and the associated colors. Subsequently, representative examples of various coloration concepts selected from rich literature are presented and discussed in view of their potentials for commercial display devices. When appropriate, other non-plasmonic coloration concepts are also included, *e.g.* structural coloration based on FP resonances and dielectric metasurfaces, to foster novel hybrid coloration concepts.

## 2. Key performance indicators

In order to evaluate the potential for plasmonic display applications, we briefly discuss the strengths and weaknesses of the respective coloration concepts in terms of key performance indicators frequently used in display technologies, such as brightness, contrast, chromaticity, resolution, angle-dependence, refresh rates and power consumption. Additionally, the lifetime, defined as the number of switching cycles without degradation, and the compatibility to commercial mass fabrication are of vital importance for potential applications and will determine the success of dynamic plasmonic colors (*19*). A more comprehensive overview on the performance indicators can be found in Ref.(*4*).

The luminance given in candela per m$^2$ is a common measure for the brightness of emissive-type and backlit-type displays. Noteworthy, the overall optical efficiency of a thin-film transistor (TFT) LC display, for example, is only about 5% because of the losses introduced by polarizers, color filters and others (*20*). For reflective displays, the brightness strongly depends on the illumination, as the ambient light is reflected by dyes and pigments. A value close to 100%, as potentially provided by strongly reflecting/scattering plasmonic nanoparticles, is highly desirable. In contrast, the typical reflectance of the electrophoretic black-and-white displays is approximately 40% (*21*). The contrast of a display is commonly defined as the difference between the on (brightest) and off (darkest) states for reflective displays and the ratio between the on and off states for emissive displays (*4*). In addition to brightness and contrast, the chromaticity is an important parameter for color printing and displays. The chromaticity characterizes how vibrant colors are. An analysis of the chromaticity is usually performed based on a 2D version of the CIE 1931 diagram by neglecting the brightness (*22*). The most saturated colors are monochromatic and are located on the edge of the CIE diagram (*23*). Every subset spanned by two or more primary colors is termed gamut. All colors within a gamut, the so called secondary colors, are mixed by combinations of the primary colors, *e.g.* implemented as sub-pixels in displays. For example, the standard red green blue (sRGB) gamut is a triangle with red, green and blue being primary colors. Another important performance indicator is the resolution given in dots per inch (dpi). Plasmonic color prints offer remarkable resolutions up to 100,000 dpi as demonstrated by Kumar *et al.* (*24*). Even though the theoretically possible maximum resolution is decreased for plasmonic color displays, if sub-pixels are integrated, it is still orders of magnitudes higher than that of commercially available displays (up to 500 dpi). Additionally, it significantly exceeds the highest resolution (~300 dpi), which can be distinguished by eye at handheld distances (*4*). The viewing angle is of vital importance as well. In the extreme case, an observer may perceive different colors, when looking from different angles. In current LCD technology, the effect is reduced by thinner liquid crystal (LC) layers. Plasmonic excitations own a strong angle dependence associated with the specific design, but commonly affect the colors strongly. Another important property of displays is the refresh rate or the switching time, respectively. Commercialized LC, LED and OLED displays offer switching times on the order of several milliseconds (*1*). In-plane switching (IPS) LCs commonly integrated as touch panels, for example, possess switching times of about 10 ms. Reflective electrophoretic displays incorporated in e-readers, for example, have significantly longer switching times in the order of 100 ms but provide an excellent bistability. The latter is accompanied by a low power consumption, since the image on the electrophoretic screen is retained even if all power sources are removed. The power consumption for emissive or backlit screens is considerably larger during operation, but typically less than 100 mW/cm$^2$ (*25*). Plasmonic color displays will have to compete with such low values as well as operational lifetimes longer than 10 years for LCDs and 3600 hours for OLED displays (*1*).

## 3. Dynamic plasmonic color generation

The realization of dynamic structural color generation is challenging, but indispensable for functional display devices. Basically, the vibrant but static plasmonic colors require a reversible modification after fabrication. Dynamic plasmonic color generation has been demonstrated utilizing various concepts. Each coloration scheme consists of two essential parts: the coloration mechanism and the coloration control. The coloration mechanism underlines how a broad range of colors are dynamically generated, *e.g.* by size variations of the nanostructures or changing the dielectric properties of the nanostructure itself or the surrounding media. It also involves the technical implementation of pixels, *e.g.* as monopixels or as sub-pixels. Ideally, one dynamic pixel composed of one or more nanostructures can exhibit any desired color. The experimental realization of such a monopixel design is challenging, since it requires plasmonic resonance shifts over the entire spectral range. In sub-pixel designs, well known from the current display technology, the perceived color of a pixel is generated by additive or subtractive mixing the plasmonic colors provided by the constituent sub-pixels. The coloration control encompasses how variations of the size or dielectric properties are experimentally realized. This is usually accomplished by a functional medium, *e.g.* electrochromic materials, liquid crystals, *etc.*, controlled by external stimuli including electric fields, light, gases, pH changes, among others.

### 3.1. Coloration mechanisms

The most straightforward coloration mechanism is to directly tune the plasmonic excitation. It is accomplished by reversibly tailoring the intrinsic properties of the nanostructure, *e.g.* the material dielectric properties, size or shape (*9*). For example, the electron density $N_e$ of a metal determines its plasma frequency $\omega_p$ (Eigenfrequency of the electron density oscillations) and thus the plasmonic resonance frequency of the metal nanostructures (Fig. 1A). As a result, the visual appearance associated with the plasmonic excitation can be directly adjusted by the metal's properties. While this appealing coloration mechanism allows for a direct color control without any additional functional material, it is usually difficult to implement for metals due to an effective Debye screening. Phase change materials, such as metal hydrides, offer a solution to it. The optical properties of magnesium (Mg) and magnesium hydride ($MgH_2$) respectively, for example, can be reversibly controlled by hydrogenation and dehydrogenation, suggesting Mg as a plasmonically active (*26*), functional material for dynamic color control (*14*, *27–29*). In analogy to plasmonic coloration, structural colors produced by dielectric metasurfaces can be actively controlled by the intrinsic optical properties, *e.g.* the absorption of the dielectric material (*30*).

Other than the intrinsic electronic properties, the size of the nanoparticle determines the perceived color as well (Fig. 1B). It is well-known that the LSPR and thereby the perceived color strongly depend on the charge distribution on the particle's surface (*31*, *32*). For small particles, the resonances are dominated by the excitation of dipolar modes. As the particle size increases, the restoring force between the opposite charges decreases and the plasmonic band appears at longer wavelengths. Thus, the size of the particle offers a direct control of plasmonic color. Additionally, if arranged in arrays, in particular closely spaced arrays, a modification of the particle size is inevitably accompanied by a change of the interparticle distances of adjacent nanoparticles. Depending on the particle separation, different effects such as near-field or far-field coupling promote a variety of coupled plasmonic modes (*33*). Since the plasmonic properties are highly sensitive to interparticle distances of closely spaced nanostructures, *e.g.* nanoparticle dimers, already minute modifications give rise to dramatic color changes. On the one hand, such coupled plasmonic systems open a pathway to continuously adjust the plasmonic color over a broad spectral range, which goes far beyond the mere size-tunability of non-interacting nanoparticles. On the other hand, the coloration based on coupled plasmonic modes comes along with significant challenges. The interparticle separation requires an excellent control with nanometer precision over the entire plasmonic color pixel, usually comprised of several nanoparticles, to ensure homogenous and vibrant colors. Basically, such dynamic distance modifications between adjacent nanoparticles as well as size changes of nanoparticles are rather difficult to achieve after fabrication. Reversible electrochemical deposition of metals onto predefined nanostructures utilizing reduction-oxidation-chemistry (redox-chemistry) offers a practical solution to size control (*34–40*), whereas mechanical strain (*41*, *42*) and configurational changes of molecules (*43*) can be employed to efficiently modulate the interparticle distances.

Resonantly excited, metal nanostructures offer strongly confined electromagnetic fields. Such highly confined near-fields dramatically increase the light matter interactions on the nanoscale, giving rise to various applications, including ultrahigh-sensitivity spectroscopy and bio-sensing, super-resolution imaging, and subwavelength optics (*44*). In refractive index-based bio-sensing, for example, variations in the polarizability of the surrounding medium alter the restoring force of the plasmon and thereby its resonance frequency (*45*). This is also a concept, which was successfully applied to dynamic plasmonic color generation. Several experimental designs were conceived, where the refractive index of a surrounding (functional) medium and hence the plasmonic excitation could be controlled by an external stimulus, such as electric fields, ions, light, electrical fields, gases and others (Fig. 1C and D). As a result, distinct plasmonic colors, predefined during fabrication, could be tuned over a broad spectral range. However, not only the real part of the complex refractive index can be modulated, but also the imaginary part (absorption). Such a controllable absorption modulates the intensity of the plasmonic excitation rather than its resonance frequency. It was successfully employed to switch distinct colors between two color states (*46*) or a color state (*e.g.* blue) and an absorbing state (black) (*47*). Remarkably, nanometer-thick layers were already sufficient to achieve full absorption due to the increased light matter interaction provided by the nanostructures.

A wide color gamut can be realized with sub-pixel designs well known from current display technologies. Such plasmonic pixels contain a certain number of sub-pixels (*e.g.* three primary colors), which are surrounded or covered by an active material. The absorption of the functional material is controlled and selectively switched, providing an individual control of each sub-pixel. Compared to monopixel designs based on mere refractive index variations, the sub-pixel design offers a substantially broader gamut, but at a price of a reduced resolution and brightness/reflection. Promising candidates for functional-medium-controlled dynamic plasmonic color generation utilizing mono- and sub-pixel schemes are electrochromic polymers and transition metal oxides, phase change materials and liquid crystals as discussed below. Moreover, liquid crystal-based approaches offer a polarization-based color selection as well as a dynamic reorientation of nanostructures embedded in the liquid crystals.

## 3.2 Coloration control

In the following, we present recently developed plasmonic coloration concepts and discuss their performance indicators in view of potential applications. A large variety of external stimuli, coloration mechanisms, coloration controls and implementations of pixels are available. In order to highlight the importance of the functional media for the dynamic plasmonic color control, we organize the review according to the functional media, *e.g.* LCs or electrochromic materials, rather than the coloration mechanism or control stimuli. For information on coloration control enabled by approaches other than functional materials or electrochemical deposition, the interested reader can refer to more general reviews (*4*, *6*, *19*).

**Hydrogenation of magnesium**

Phase transition metal hydrides are representative candidates for dynamic plasmonic color generation. Induced by the absorption/desorption of hydrogen, metal hydrides alter their crystallographic and electronic structure, resulting in dramatic changes of the optical properties along with the metal to insulator transition. Among a variety of phase transition metal hydrides, Mg and $MgH_2$ respectively, have attracted particular interest in the field of dynamic plasmonic color generation due their unique optical and phase transition properties

(*26*). First, Mg exhibits an excellent plasmonic response in the visible regime, compared to other common phase transition metal hydrides such as palladium (Pd) or yttrium (Y), and the widely used passive metals like Au, Ag, or Al. Second, its optical properties can be reversibly switched between metal (Mg) and dielectric ($MgH_2$) states upon hydrogen loading/unloading (*48*). Compared to other functional materials employed for active plasmonics, such as $VO_2$, Ga, germanium-antimony-tellurium (GST), perovskites or graphene, Mg additionally offers a much larger spectral tuning range and contrast.

As shown in Fig. 2A, Duan *et al.* demonstrated dynamic plasmonic color generation utilizing catalytic Mg metasurfaces (*27*). By hydrogen exposure, the constituent metallic Mg nanoparticles were transformed to dielectric $MgH_2$ particles and the plasmonic colors were erased. The dynamic plasmonic pixels comprised of Mg nanoparticles sandwiched between a Ti/Pd capping layer and a titanium (Ti) buffer layer were fabricated by electron beam lithography (EBL) and evaporation techniques. Depending on the geometrical configuration, *e.g.* particle sizes and periodic interparticle distances, vivid reflective colors emerged from LSPRs and Rayleigh-Wood anomalies. As an example for dynamic color control, the authors fabricated a high-quality plasmonic microprint of the Max-Planck Society's Minerva logo. When exposed to molecular hydrogen in a specially designed gas chamber, the Pd capping layer catalyzed the dissociation of hydrogen molecules into hydrogen atoms, which then could easily diffuse into the Mg nanoparticles. The incorporation of hydrogen into the metallic Mg (hydrogenation) led to the formation of dielectric $MgH_2$, which was accompanied by a large volume expansion. The reversible transformation from the metal to dielectric state gave rise to a series of color changes, until all colors vanished after roughly 10 minutes. The gradual color change was related to a decrease of the metallic fraction of the nanoparticles, the formation of $MgH_2$ surroundings and an expansion of the Mg particle volume. If exposed to oxygen (dehydrogenation), dielectric $MgH_2$ was transformed to $H_2O$ and metallic Mg. The presence of oxygen avoided the buildup of hydrogen at the Pd surface and thus facilitated the desorption of hydrogen from the $MgH_2$. As a result, the logo was gradually restored to its initial color state within a few seconds. The hydrogenation/dehydrogenation process was reversible and hence the colors could be restored closely to their initial states, after more than 10 cycles. Remarkably, by stopping the hydrogen or oxidation exposure, the microprint could be 'frozen' at any intermediated state between the fully dielectric and the fully metal states. Such a passive and long term stability of a particular color state is highly desired for display devices, *e.g.* electronic papers, since it lowers the display's power consumption.

More brilliant and saturated structural colors were achieved by the integration of pixelated Fabry-Perot cavities comprising Mg elements as shown in Fig. 2B (*14*). A FP cavity is an optical resonator typically composed of two facing mirrors and a dielectric material placed in between. Depending on the dielectric thickness and its optical properties, the light field is selectively enhanced through resonance. Such FP cavities are popularly applied in the field of optics, *e.g.* as optical filters. By using grayscale electron beam lithography, color pixels (500 × 500 nm$^2$) comprised of FP cavities with various heights have been fabricated. Each cavity was formed by a dielectric hydrogen silsesquioxane (HSQ) pillar sandwiched between an aluminum (Al) mirror and a metallic capping layer composed of Mg/Ti/Pd (50 nm/2 nm/3 nm). The key element for dynamic control of the reflected light was the functional Mg/Ti/Pd capping layer, which could be reversibly switched between a reflective metal state and a nearly transparent state upon hydrogenation and dehydrogenation. In its metal state, the thick Mg/Ti/Pd capping layer efficiently reflected the visible light and no cavity modes could be excited (blank state). Upon hydrogen exposure, the Mg/Ti/Pd multi-layer gradually was transformed to MgH$_2$/TiH$_2$/PdH and the effective thickness of the capping layer was significantly reduced. Accordingly, the effective height of the dielectric spacer, given by the HSQ pillar height and the MgH$_2$ thickness, increased. Consequently, light passing through the thin lossy metallic TiH$_2$/PdH capping layer excited cavity resonances. Depending on the particular height of the pillars predefined in the fabrication process, the cavity resonance gave rise to various vivid colors (color state). Due to the well-modulated and sharp FP resonances, the generated structural colors were more brilliant, more saturated, and richer than those originating from Mg nanoparticle pixels.

The Mg-based coloration concepts presented so far are not capable of selective and local switching of the plasmonic pixels via hydrogenation. As shown in Fig. 2C, Duan *et al.* addressed this important issue and proposed a scanning plasmonic color display operating in a reflection mode (*28*). Following this concept, subwavelength plasmonic pixels were laterally controlled by employing an underlying Mg layer as a scanning screen. The generated plasmonic colors were laterally erased or restored, when the Mg layer was reversibly transformed between a metal and dielectric state upon (de)hydrogenation. To this end, periodically arranged Al nanoparticles were fabricated by electron beam lithography on a top of a 20 nm thick $Al_2O_3$ layer residing on the Mg scanning screen. An additional Pd strip placed at a particular edge of the screen served as a gate for hydrogen loading and unloading. In the non-hydrogenated state the particle-on-a-mirror geometry provided a wealth of plasmonic colors determined by the particular geometric configuration. During hydrogenation, the hydrogen could only enter the Mg via the Pd gate, which catalyzed the dissociation of molecular hydrogen into atomic hydrogen. Consequently, the absorption of hydrogen started from the Pd gate and uniformly evolved along a certain direction with a speed limited by the diffusion of hydrogen in Mg and $MgH_2$, respectively. The accompanied metal to dielectric phase transition led to the formation of $MgH_2$ and thereby a dynamic erasure of the brilliant colors. This process was reversible through dehydrogenation using oxygen. The dehydrogenation also started at the Pd gate and laterally propagated along the hydrogen diffusion routes leading to the restoration of the initial metallic Mg screen and the associated plasmonic colors. By tailoring the position, number, and geometry of the Pd gates, this scheme could be extended to generate a variety of scanning effects and applications, such as multifunctional animations or information encryption.

One major drawback of the previously discussed coloration concepts is the need for cumbersome gas chambers required for (de)hydrogenation, which hamper practical real-word applications. Huang *et al.* overcame this limitation and took the Mg-based coloration concept an important step forward. By integrating a nanoscale solid-state proton source into Mg-based plasmonic devices, the optical properties of Mg could be selectively and locally adjusted (*29*). As depicted in Fig. 2D, the electrically switchable plasmonic device was comprised of multi-layer system of Al(100 nm)/Mg(40 nm)/Pd(5 nm), periodically arranged Al nanodisks embedded in a 40 nm thick gadolinium oxide ($GdO_x$) layer, and a 3 nm thin gold (Au) layer. In this configuration, Mg served as a switchable mirror and gold as the top gate electrode. The particle-on-a-mirror geometry provided a variety of reflective plasmonic colors, mainly determined by the diameter and separation of the nanodisks as well as the spacing between the reflective mirror and the Al nanodisk arrays. When 5 V were applied for 120 s, water molecules from moisture near the $GdO_x$/Au interface were split into molecular oxygen ($O_2$) and hydrogen ions ($H^+$). The gate bias then drove the extracted protons through the proton conducting $GdO_x$ layer towards the bottom electrode (Al/Mg mirror) and the Mg layer was loaded with hydrogen. Remarkably, the hydrogenation could be selectively and locally controlled by particular nano- and/or micro-structured gold electrodes serving as sources for hydrogen ions. As a result, the previously metallic Mg was switched to transparent dielectric MgH$_x$ and the Al layer served as the new bottom mirror. The increased effective thickness of the dielectric spacer, given by the thickness of MgH$_x$ and GdO$_x$, caused a blue-shift of the plasmonic resonance and a change of the plasmonic colors. When -2 V were applied for one hour, the plasmonic colors were restored closely to their initial states demonstrating a good reversibility, even after hundreds of cycles. The switching time between the color states was basically determined by three processes, which were the water hydrolysis reaction, the transport of protons through the GdO$_x$ and the (de)hydrogenation of the switchable Mg/MgH$_2$ mirror. While the hydrolysis and proton transport were rather fast processes (~10 ms), loading and particularly unloading of hydrogen slowed down the overall switching speed significantly. Due to the crucial importance of the switching time for display screens, the authors suggested Mg-free designs, where mere refractive index changes of GdO$_x$ were exploited to switch colors generated by thin film interferences.

Evidently, Mg-based coloration can be considered as an interesting candidate for dynamic color generation because of its ability to produce tunable vivid colors. Nevertheless, in the current stage practical applications are limited by the long response time, complex and cost-intensive fabrication, wear-and-tear of the Mg/MgH$_2$ elements during phase change, low durability, gas cells and lack of local and selective addressability on the pixel level. Whereas the latter could be significantly improved by nanoscale proton pumps as demonstrated by Huang *et al.*, the intrinsically slow switching time remains a major limiting factor. Due to that restriction, Mg-based dynamic color generation seems rather more suitable for applications in information encryption, anti-counterfeiting or other hydrogen-related areas.

**Electrochemical deposition of metals**

Electrochemical deposition allows for a precise and controlled growth of nanometer thick layers of metals and conducting metal oxides on surfaces and nanostructures (*49*). Usually, electrochemical cells employed for deposition are composed of a pair of electrodes and metal ions in a reducible form dissolved in an electrolyte between them. By applying a voltage, metal ions are reduced and deposited quasi-instantaneously onto an electrode, *e.g.* onto metal nanoparticles. When an inverse voltage is applied, the previously deposited metals can be oxidized and dissolved again.

In 2012, Araki *et al.* applied the electrochemical deposition concept to dynamic plasmonic color generation (*34*). The authors demonstrated an electrochromic device comprised of a gel electrolyte containing silver ions ($Ag^+$), which was sandwiched between two facing electrodes, a flat indium tin oxide (ITO) electrode and an ITO particle-modified electrode. In the initial state, the device was transparent. When Ag was deposited on the particle-modified electrode, the cell appeared black due to multiple scattering and/or absorption of the light by the aggregated Ag particles. The deposition of Ag on the flat surface created a silver mirror leading to a specular reflection. Later, the same group extended the concept and produced red and blue colors in addition to the black, mirror and transparent states by controlling the nucleation and growth at different voltages (*35*). Similar to the basic configuration, the dimethyl sulfoxide-based gel electrolyte was sandwiched between the electrodes and contained silver nitrate, copper chloride and tetrabutylammonium bromide (Fig. 3A). When an appropriate constant voltage was applied, Ag was continuously deposited on the particle-modified or flat ITO electrode, resulting in a black state or mirror state, respectively. To generate blue and red colors, two different voltages $V_1$ and $V_2$ were utilized successively. The first voltage $V_1$, with values more negative than the critical nucleation voltage, was applied very briefly to initiate Ag nucleation at the flat ITO electrode. Immediately afterwards, a second voltage $V_2$, with absolute values smaller than the nucleation voltage, was applied promoting the growth of the Ag nuclei. As the particles started to grow, a plasmonic band was formed causing a color change of the cell's transmittance from transparent to red within a few seconds. With increasing particle sizes (corresponding to longer deposition times), the LSPR shifted towards 600 nm, resulting in a blue color. Additionally, an increased coupling between neighboring nanoparticles further red-shifted the plasmonic resonance. Even though a very uniform growth of the nanoparticles could be measured, the standard deviation in size and the multiple interactions between the randomly distributed adjacent particles resulted inrelatively broad absorption peaks (compared to a solution of chemically synthesized Ag nanoparticles), which limited the purity of the colors. If positive voltages were applied, the electrodeposited Ag was oxidized and dissolved resulting in an increased transmittance of the electrochemical cell. Following this electrochemical approach, five different color states including black, mirror, transparent, red and blue could be reversible switched by controlling the voltage between the ITO electrodes.

Continuous color tuning from yellow to blue was achieved by reversibly electrodepositing and dissolving Ag shells surrounding Au nanoparticles (*36*). Using an anodized aluminum oxide (AAO) template as an etching mask, the authors etched periodic arrays of holes into a 50 nm thick $SiO_2$ layer residing on a conductive ITO substrate (Fig. 3B). Thermal evaporation of gold and subsequent partial removal of the gold covered AAO resulted in the deposition of Au nanoparticles (diameter 40 nm) in wells. Integrated in a functional electrochemical cell, the nanoparticles served as the cathode during silver deposition (reduction process). The facing counter electrode was a 1 nm-thick platinum (Pt) layer deposited on a glass window. When a voltage was applied, Ag was electrochemically deposited from an electrolyte containing silver nitrate as well as potassium nitrate. A silver shell started to grow on the Au nanoparticles. The particular size of the Au-Ag core shell particles, and thus their plasmonic colors, could be efficiently controlled by the deposition times. At zero deposition time, the plasmonic excitation of the pure Au nanoparticle array appeared at 580 nm and a yellow color was observed in the cell's reflectance. As the silver shell increased, the plasmonic band blue-shifted towards 400 nm, which is a characteristic frequency for LSPRs of pure Ag nanoparticles. Intermediated plasmonic states and related colors during the metal deposition were determined by the Ag shell thicknesses (material factor) and the shape of the Au-Ag core shell particles (shape factor). For longest deposition times (50 s), a saturation of the blue shift was expected, but a red-shift was observed. According to the authors, the transformation from hemispherical to prolate hemispherical particles (shape factor) during the growth process contributed to the non-monotonic relation between the deposition time and resonance frequency (color). Other works found that the material factor and shape factor had the same peak shift-directions for the presented Au-Ag core-shell configuration, considerably helping to achieve a wide tuning range (*37*). The respective color state could also be restored closely to initial yellow color, which is a prerequisite for reversible color switching. In that case, the functionality of the Pt electrode and Au nanostructure electrode were exchanged: Pt served as the cathode and the Au nanostructures as the anode. Upon applying a certain voltage for 60 s, oxidation took place at the gold electrode, the Ag shell was dissolved, and the pure Au nanoparticles remained. The switching between the initial color and the color state was shown for 10 cycles with a fairly good quality. However, deposition of Ag on the Pt electrode window lowered its transmittance and thus reduced the brightness of the generated color significantly. Additionally, parasitic growth of Au/Ag alloys outside of the Au nanoparticles blue-shifted the color of the device already after the third switching cycle. Both effects make potential devices rather inapplicable for real world applications, where a high color durability is desired.

A significantly increased endurance of 200 cycles of operation with a good color durability was reported by Wang *et al.* (*37*). Similar to the previously discussed study, the authors fabricated gold nanoparticles in hexagonal arranged wells, packaged them into an electrochemical cell and filled the device with a dimethyl sulfoxide-based gel electrolyte containing $Ag^+$ ions. By applying different voltages, Ag shells of different thicknesses were reversibly electrodeposited and removed onto/from the Au nanostructures. As a result, the plasmonic reflection of the Au-Ag core shell particles could be continuously tuned between 430 nm (blue color) and 650 nm (red color) within a few seconds. Moreover, distinct colors could be electrically actuated during the deposition and stripping process, offering many potential applications, for instance, active camouflage, among others. As an example, the authors constructed a mechanical plasmonic chameleon that rapidly adapted the surrounding of an object (Fig. 3C). The mechanical chameleon was covered with scale-like plasmonic color patches (typical size of up to several $cm^2$) comprised of the aforementioned electrochemical cells. Additionally, two sensors detecting the surrounding color were integrated as well. The sensor output was analyzed by a microcontroller, which subsequently applied a voltage of 1.5 V for a suitable time to the corresponding plasmonic patch. As a result, the color matching the background was displayed.

Noteworthy, the key parameter for size-based LSPR tuning is the optical size (*e.g.* optical length) of the nanostructure and not its geometrical size (*e.g.* the geometric length). Nanoparticles composed of two segments, *e.g.* a metal and dielectric section with length $L_m$ and $L_d$, have an overall geometrical size of $L_m + L_d$, but the LSPR is mainly related to the length $L_m$ of the metal nanostructure. Recently, this fact was exploited for LSPR tuning enabled by reversible oxidization (reduction) of a copper (Cu) film containing nanoslits (*38*). In this configuration, Cu oxidized preferably at the nanoslits' interfaces and thus modulated the optical lengths, whereas the geometrical lengths remained constant. LSPR shifts of up to 200 nm were observed in the reflectance spectra. The concept was also applied to gold/silver (chloride) core-shell nanostructures, where the shell could be reversibly switched between semiconducting silver chloride and conductive Ag upon redox-chemistry (*39*). However, both studies only comprised extensive analysis of the resonance shifts during redox reactions but did not demonstrate dynamic color generation.

In summary, reversible electrochemical deposition has been successfully applied to dynamically generate plasmonic colors, covering the full visible spectral range. The demonstrated devices offer fairly good brightness. The low purity of colors caused by the broad and overlapping resonances can be improved by using more sophisticated arrangements of nanostructures providing spectrally sharp Fano-resonances, for example. However, the coloration times on the order of several seconds hamper potential applications for color displays with high refresh rates. Additionally, the color durability/stability over many cycles of operation suffers from Ag oxidation, wear-and-tear (due to cyclic Ag deposition/removal), and parasitic deposition of metals. Recently, it has been shown that the parasitic random nucleation can be improved by using hollow shells of Au/Ag alloys as stable anchoring sides (*40*). However, wear-and-tear is of particular importance for strongly interacting nanoparticles, where slight distance modifications result in large spectral shifts and related color changes. Even though the fabrication and operation can be easily scaled up, the implementation of electrochemical plasmonic cells as functional plasmonic pixels in color displays demands more engineering. Particularly, the addressability and fabrication of micrometer or even nanometer-sized pixels composed of electrochemically-switchable Au-Ag core-shell nanoparticles appears as a daunting challenge.

**Liquid crystals**

Inspired by the LCD technology, many research efforts in the field of dynamic plasmonic color generation have been devoted to LC-based coloration control. LCs are an intermediate state between a solid and a liquid with unique properties, enabling a wealth of applications in today's world (*50*, *51*). In general, the orientation of the constituent LC molecules or birefringence can be optically, thermally or electrically switched on millisecond timescales for an infinitely number of times without any quality deterioration. LCs as a functional optical material provide three mechanisms to dynamically control the plasmonic colors. First, LCs rotate the polarization state of light incident onto or scattered from anisotropic nanostructures due to their birefringent nature. Combined with an additional polarizing element, distinct polarization states of the plasmonic excitation and the associated colors can be selected. Second, the anisotropic refractive index of LCs surrounding the metallic nanostructures can be electrically or optically controlled. As elaborated above, variations in the surrounding refractive index change the plasmonic resonance frequency and thereby the perceived color. Third, anisotropic gold nanoparticles dispersed in a LC host can be aligned by an electrical field. An additional polarizer selects the desired polarization state and plasmonic color, respectively.

In analogy to the standard LCD technology, several groups employed LCs as polarization rotators in different configurations (*46*, *47*, *52–55*). For example, Olson *et al.* used LCs as a polarization rotator to electrically switch plasmonic pixels from a color to a black state as schematically shown in Fig. 4A (*47*). The vivid structural colors were generated by a hexagonally arranged Al nanorod array fabricated using EBL and e-beam evaporation. When illuminated by light in an attenuated-total-reflection-type geometry, lattice modes (tunable via the array's periodicity) as well as polarization-dependent LSPRs (tunable via the geometric dimensions of the Al nanorod) were excited. The interaction between the broadband LSPR and the spectrally sharp lattice resonance gave rise to a Fano-type-interference. Both the Fano-type resonance and the far-field diffractive coupling narrowed the plasmonic resonances substantially. This led to extremely vivid colors adjustable across a wide spectral range. Remarkably, the LSPRs and the associated coloration were highly polarization-dependent, since only light polarized parallel to the long rod axis was scattered efficiently. Square pixels (side length of 1.5 mm) composed of such polarization-selective plasmonic nanorods were incorporated into LCs to demonstrate on and off switching of red, green and blue colors. To this end, a 6µm thick layer of the LC 4-cyano-4′pentylbiphenyl (5CB) was sandwiched between polyimide-covered Al nanostructures on ITO and a facing polyimide-covered ITO substrate without nanostructures. In the voltage-off state, the LCs were aligned in their twisted nematic (TN) configuration, causing a 90° rotation of the light scattered by the rod-like nanostructures. The scattered and rotated light passed through a polarization filter installed before the detector and the pixels were visible. When 20 V were applied, the LCs aligned parallel to the applied electrical field across the LC cell. The LCs did not modulate the polarization and the polarization imparted by the rod-like nanostructure remained. As a result, the polarizer blocked the light scattered from the plasmonic pixels and the pixels appeared black in the voltage-on state. In view of potential applications, this approach is appealing because of its wide color gamut, but has two significant conceptual drawbacks in addition to its high power consumption and lack of compactness. First, the diffractive-based plasmonic colors produced in this study strongly varied with the viewing angle. The use of additional diffusors might mitigate the angle dependence but decrease the brightness. Second, the efficiency defined as the ratio between the scattered and the incident light intensities was below 4%. This low value was mainly due to the relatively large inter-particle distances, which could not be increased without changing the plasmonic colors.

Lee *et al.* took the concept one step further and electrically switched a single plasmonic pixel between two primary colors instead of simply on and off (*46*). Thereby, a single multicolor pixel could be realized by mixing the two primary colors. To this end, rectangular lattice nanohole arrays with asymmetric periodicities in *x*- and *y*- directions providing polarization-dependent extraordinary optical transmission (EOT) were fabricated with EBL (Fig. 4B). Between the nanohole array and a facing ITO-coated glass substrate, a 5 μm thick layer of electrically switchable TN LC molecules was integrated to modulate the polarization of the incident light. An additional polarizing layer transmitting *y*-polarized light only was added on top of the glass substrate, making the device highly compact compared to the previously presented design. At the electrically saturated state (5 V), the LC molecules were aligned vertically, and the polarization predefined by the polarizing layer was not affected by the LC molecules. Such linearly *y*-polarized incident light passed through the LC layer and excited the hybrid modes of LSPRs and SPPs, giving rise to EOT up to 40%. The associated colors could be adjusted over a broad spectral range by altering the structural conditions, such as the hole size, hole shape, lattice periodicity, film thickness and dielectric environment. Figure 4B shows a palette of vivid colors generated by various periodicities and hole sizes. At the off state, the TN-state, the linearly *y*-polarized incident light was rotated to the *x*-polarization by the LC layer. Because of the different lattice periodicities in the *x*- and *y*- directions, the plasmonic resonance and, hence, the associated color changed significantly, *e.g.* from orange to green as shown in Fig. 4B. As expected for LC-based devices, the coloration time was on the millisecond scale (down to 7 ms). Additionally, the authors applied intermediated voltages and thus actuated various LC states, which only partially rotated the polarization of the incident light. Consequently, the two primary colors determined by the plasmonic resonances in the *x*- and *y*-directions could be superimposed and various secondary colors were generated. As a result, only one pixel composed of the minimum number of holes necessary to provide EOT, delivered various secondary colors. However, the realization of a wider gamut requires a broader tuning range or the implementation of sub-pixels.

A promising strategy to realize a wider gamut utilizing one single plasmonic pixel is to reversibly adjust the anisotropic dielectric properties of the surrounding LCs. The refractive index of LCs is anisotropic and can be reversibly switched between an ordinary

($n_o$) state and an extraordinary ($n_e$) state by phase transitions or electrical field alignment of the LC molecules. The changes can be introduced either electrically, thermally or even optically. Naturally, optical approaches are appealing, given their remote, temporal and spatial control. Already in 2012, Liu *et al.* demonstrated all-optical control of LCs for dynamic plasmonic coloration (*56*). The authors fabricated plasmonic transmission filters composed of annular aperture arrays with focused ion beam (FIB) lithography and covered them with a layer of a photoresponsive LC, 4-butyl-4-methy- oxyazobenzene (Fig. 4C). The azobenze derivative was switched from a *trans*-isomer (rod-like shape) to *cis*-isomer (bent shape) state upon UV illumination. The bent shape of the photochromic LCs, disrupted the local order (nematic phase) and an isotropic LC phase was formed. As a result of the LC phase transition, the refractive index decreased by 2.5% and modulated the transmission properties of the plasmonic nanostructures by more than 40% at resonance. Upon illumination by visible light or thermal isomerisation, the photochromic LC returned from the *cis*-state to the *trans*-state and the nematic phase with its specific refractive index was restored. While this approach employed metal annular aperture arrays, providing planar surface plasmons and cylindrical surface plasmons for coloration, refractive index modulation of LCs has also been applied to switch structural colors generated by dielectric metasurfaces (*57*).

Apart from optical control, electrical switching of the LCs' anisotropic refractive index has been applied to dynamically generate plasmonic colors as well. During the last few years, several proof of concept experiments have been reported, all limited by a small color tuning range caused by the modest shifts of the plasmonic excitations. Recently, the concept has been significantly advanced (*58*). A continuous color tuning over a considerably large spectral range (95 nm) was achieved by utilizing highly birefringent LCs directly coated on arrays of shallow Al nanowells (Fig. 4D). The key to a wide color tuning was a shallow design of the plasmonic nanowells, which allowed for a complete LC reorientation and, concurrently, an optimized refractive index change. Such shallow nanowells were imprinted on a polymer and covered with a smooth Al overlayer (30 nm). The master patterns fabricated by direct laser writing did not only allow for highly scalable fabrication but also for nanoimprints on flexible substrates. A 5 µm thick layer of highly birefringent LC (Hi-Bi LC) residing on top of the nanowells and a facing ITO-covered superstrate completed the compact LC cell. Light impinging on the Al nanowell arrays excited grating-coupled SPPs and the reflected light, typically 50% to 80%, determined the perceived color. As a matter of fact, the SPPs and the associated colors were highly depended on the arrays' periodicity and the refractive index of the surrounding LC. The latter could be continuously adjusted between $n_o = 1.55$ and $n_e = 1.97$ by the LCs' orientation controlled by a voltage applied across the LC cell. At the off state, a shallow design of the nanowells allowed for a homogeneous alignment of the LCs parallel to the Al surface inside the nanowells. When an external voltage was applied, the LC molecules started to vertically rotate from their initial state, until they aligned along the electrical field inside and outside the wells. The highly reversible orientation process was accomplished typically in less than 90 ms. Other designs, *e.g.* nanowells with larger depth-to-diameter ratio, suffered from a vertically alignment already at the off state, as a result of the extremely high surface sensitivity of the LC orientation. In contrast to such non-optimized designs, the shallow nanowells allowed for a complete reorientation, resulting in a maximum refractive index change of 0.4 and thereby SPP resonance shifts up to 95 nm. Based on this concept, Franklin *et al.* fabricated a plasmonic color micro image of the Afghan Girl, which could be electrically tuned. Remarkably, relatively high electrical fields were applied to switch all colors successfully. The authors attributed this fact to the surface anchoring forces of the LC molecules, which were substantially higher in the vicinity of the nanoparticles compared to the bulk of the LC. Furthermore, the forces strongly depended on the topography of the Al surface, *e.g.* the periodicity of the nanowell arrays, giving rise to different saturation voltages for green (2.5 V $\mu m^{-1}$) and pink (10 V $\mu m^{-1}$) colors. While this grating-based coloration mechanism has intrinsic restrictions with respect to minimum pixel size (approximately 2.1 µm) and viewing angle (invariant up to 20° only), the incorporation of LCs provided vivid colors tunable over a wide spectral range. This basically offers the possibility to design pixelated color displays based on two tunable color sub-pixels instead of three static color sub-pixels (*e.g.* RGB) as commonly integrated in today's displays.

A very important step towards single plasmonic pixels producing the full RGB color basis set was taken by the same group in a subsequent study (*54*). The authors significantly advanced the basic concept and exploited the polarization rotation in addition to the anisotropic refractive index. By the combination of both LC effects, the reflective color of a single plasmonic pixel could be dynamically tuned over the full visible spectral range as a function of the applied voltage. Similar to the basic design, the LC cell contained shallow nanowells, LCs as a functional medium, and an ITO-covered electrode (Fig. 4E). However, instead of a smooth Al film, a rough Al layer (30 nm grain diameter) was deposited on the shallow nanowells. The presence of an anisotropic refractive index medium near the rough Al surface induced a polarization dependence on the grating-coupled SPPs. Consequently, at the off state, two primary colors, *e.g.* blue and red, could be generated, when illuminated by light polarized parallel or perpendicular to the LC director near the nanostructures, respectively. At voltages below 3.5 V, the orientation of the LCs near the nanostructures (surface LCs) remained the same, but the bulk LCs realigned and the polarization of the incident light was rotated accordingly. Depending on the polarization state of the incident light, which was directly adjustable via the applied voltage, different secondary colors with distinct polarizations were realized by mixing the primary colors red and blue. At voltages higher than 3.5 V, the orientation of the bulk LCs saturated and the surface LCs started to realign vertically as elucidated for the basic design. The reorientation of the surface LCs was accompanied by an increased refractive index, leading to a red-shift of the SPPs and an associated green color. At saturation voltages (50 V), both bulk and surface LCs, were completely aligned vertically, resulting in a loss of the birefringence and polarization dependence of the Al nanostructured surface. In order to demonstrate the approach's capability for an easy implementation in current display devices, micrometer-sized plasmonic pixels were integrated in a conventional transmissive TN LCD panel. Also at this early non-optimized state, the authors could demonstrate switching times below 70 ms. Even though the angle invariance and the generation of black and grey colors require further engineering efforts, the approach holds great potential. In particular, the ability of producing the RGB color basic set with one pixel only, may lead to a further improvement of the already excellent resolution without reducing the (sub-)pixel size.

Furthermore, plasmonic colors were dynamically controlled utilizing anisotropic metal nanostructures embedded in a LC host. Nematic dispersions of nanoparticles, such as nanorods, nanoplatelets *etc*., were obtained by attaching LCs molecules to metal nanostructures utilizing special functionalization chemistry (*59–63*). The anisotropic nanoparticles followed the director **n** describing the average local LC molecular orientation, which could be adjusted by an applied electrical field (*62, 63*). For example, Liu *et al.* integrated gold nanorods aligned parallel to the director into a LC host as schematically shown in Fig. 4F (*62*). If illuminated by white light polarized along the director **n**, longitudinal LSPRs were excited and an associated green color was transmitted. At the on state (4 V), the director as well as the nanorod rotated mechanically and eventually aligned vertically. Upon light illumination, the transverse LSPR was excited causing a red color appearance. As expected for LCs, the switching process was reversible, however, the switching took hundreds of milliseconds, which is considerably longer compared to other LC approaches. Even though millimeter-sized plasmonic images could be switched between two colors, the method has substantial restrictions, in particular, for the implementation and actuation of micrometer-sized pixels or even sub-pixels. This also holds true for other reorientation-based approaches, including optically switchable gold nanorods embedded in organic suspensions (*64*) or magnetically switchable nanoparticles in ferrofluids (*65*).

The specific performance and thus the applicability to display technology crucially depends on the respective LC approach. In view of a possible integration in pixelated display devices, the mechanical rotation of nanostructures embedded in nematic LCs seems rather ineligible. Coloration control based on mere modulation of the anisotropic LC refractive index has a certain potential, but only parts of the visible spectral range are covered generally. Approaches utilizing LCs as polarization rotators are promising for plasmonic displays, since they potentially provide a wide gamut, if sub-pixels designs are implemented. The most promising LC-based coloration control combines both the refractive index modulation and polarization rotation. In this combination, the number of sub-pixels is significantly reduced to even a single pixel, potentially increasing the resolution without reducing the pixel size. All the presented LC-based coloration approaches offer faster switching times, when compared to the aforementioned hydrogen and electrochemical deposition-based coloration controls. Additionally, the already existing knowledge on the LC technology facilitates the integration of LCs into plasmonic nanostructure-based systems. The use of LCs to control plasmonic colors comes along with the shortcomings already known from the commercial LCD display technology. First, a constant power is required to hold the current color state. Second, micrometer-thick LC cells and additional polarizers for color filtering lower the reflection efficiency and increase the fabrication complexity. More research endeavors need to be exerted to address these issues.

**Electrochromic polymers**

Electrochromic (EC) polymers, such as polypyrrole, polythiophene, polyaniline (PANI), and poly(3,4-ethylenedioxythiophene) (PEDOT), offer reversible color changes during electrically controlled redox-reactions. During the reduction (gain of electrons) or oxidation process (loss of electrons), the electronic properties of the conductive (or pi-conjugated) polymers are altered considerably, resulting in strong variations of the complex dielectric function and thereby the absorption colors (*66*). The excellent reversible and continuous control over the redox states suggest conductive electrochromic polymers as promising candidates for display applications (*5*). In particular, their vibrant colors, low power consumption, and readily processability at low costs - even on flexible substrates - are appealing for full-color reflective displays, *e.g.* electronic papers or electronic billboards. Nevertheless, substantial shortcomings still hamper their implementation in commercial display devices. Most importantly, relatively thick polymer layers are essential to generate a sufficiently high contrast. Such thick layers, however, limit the ionic diffusion in the electrochromic polymer, resulting in considerably long switching times (*67*). Typical switching times $\tau \propto L^2/D$, with the ionic diffusivity $D$ and layer thickness $L$, between different color states are on the order of seconds (*68*). In addition to the need of thick layers, a substantial contrast requires relatively high electrochemical potentials during the switching process, which deteriorate the polymers. Finally, a single type of electrochemical polymer cannot cover the entire visible range. In principle, the latter challenge can be addressed by subtractive (cyan, magenta and yellow) or additive (red, green and blue) color mixing utilizing sub-pixels. However, sub-pixels require additional conjugated polymers (at least three) and spacer layers, which increase the fabrication complexity and further reduce the switching speed as well as the reflection efficiency. Hybrid systems of electrochromic polymers and plasmonic nanostructures can help to overcome these limitations. Nanometer thin layers of electrochromic polymers located in the plasmonic hot spots basically fulfill the same purpose as largely extended electrochromic polymer layers. Consequently, the switching time is improved by orders of magnitudes, offering higher contrasts and less deterioration because of lower switching voltages. Additionally, the plasmonic elements widen the gamut, if sub-pixel approaches or single plasmonic pixels providing widely tunable plasmonic excitations are employed.

Following this approach, Xu *et al.* demonstrated high contrast full-color electrochromic switching utilizing plasmonic nanoslit arrays as transmission color filters (*68*). The deep sub-wavelength nanoslits (width 70 nm) arranged in arrays with different periodicities (from 240 nm to 390 nm) were fabricated by focused ion beam milling in a 250 nm thick Al electrode (see Fig. 5A). An additional $Si_3N_4$ layer was added beneath the Al electrode to narrow the plasmonic linewidth and further improve the plasmonic color purity. The nanostructured electrode, a Pt counter and a reference electrode were immersed in an electrochemical cell and a 15 nm thick layer of electrically switchable poly(2,2-dimethyl-3,4 propylenedioxythiophene) (PPDOT-Me$_2$) was electrodeposited onto the nanoslit arrays subsequently. By applying certain low voltages to the nanostructured electrode, electrons from the metal and ions from the electrolyte were injected into (reduction) or removed from (oxidation) the conjugated polymer, respectively. At the oxidized state ("OFF"), PPDOT-Me2 exhibited a broadband optical absorption peak in the center of the visible spectral range, which resulted in a uniformly black color. At the reduced ("ON") state, PPDOT-Me2 was transparent. Incident light thus passed through the EC polymer and excited coupled plasmonic modes that propagated along the nanoslits and Al surfaces, giving rise to distinct plasmonic colors. Depending on the slits' periodicity predefined during nanofabrication, distinct vibrant colors covering the entire visible spectral range could be generated (see Fig. 5A). The switching contrast between the color and black states was 73-90% and the absolute transmission efficiency at the filtered wavelengths ranged from 13 to 18%, which is fairly lower than that of commercial LCD pixels. Basically, more efficient polymers and optimized plasmonic nanostructures should provide higher transmittance efficiencies and switching contrasts. As expected, the implementation of thin EC polymer layers significantly reduced the switching times. Even faster switching speeds below 10 ms, which are already comparable to commercially available displays, were achieved with a modified plasmonic nanoslit design covered by PANI instead of PPDOT-Me$_2$. Additionally, gold was used instead of aluminium and no $Si_3N_4$ waveguide was added beneath the nanoslits. However, the device only operated in the red spectral range due to the high SPP propagation losses of the incorporated Au elements. According to the authors, the switching times could be further reduced by optimization of the electrochemical cell, *e.g.* a decrease of the cell's capacitance and the electrolyte's resistance. Additionally, a low power consumption (voltages up to 0.6 V only) and a good durability without polymer degradation after thousands of cycles make this approach promising for plasmonic video displays with high refresh rates.

Another EC-based plasmonic coloration concept with potential applications for flexible plasmonic video displays and full color electronic papers has been suggested by Xiong *et al.* (*69*).The authors electropolymerized thin layers of polypyrrole with thicknesses between 110 and 260 nm on a nanohole array residing on an $Al_2O_3$ layer supported by Ag mirror (see Fig. 5B). The thickness of the $Al_2O_3$ spacer layer was varied between 40 nm and 95 nm to generate various structural colors originating from cavity resonances. Strong resonant scattering and excitation of surface plasmons by the nanohole arrays further enhanced the coloration efficiency. Immersed in a water-based electrolyte containing

NaDBS and LiCl, polypyrrole coated on the metal-insulator-nanohole (MIN) film could be electrically switched between the reduced and the oxidized states by applying appropriate voltages. As for other conductive polymers, the injection and extraction of electrons and ions from the polymer altered its charge state and thus its absorption properties. In the conductive state ("Off"), polypyrrole strongly absorbed the visible light, resulting in a uniform black color. When polypyrrole was reduced to its neutral state ("On"), the band gap was modified and, concurrently, its absorption was lowered. The incident light could then pass through the EC layer and excited SPPs as well as cavity resonances. As a result, distinct colors predefined by the $Al_2O_3$ layer thickness were reflected with a high efficiency (reflection > 90%). The high-contrast switching between the color and black states took hundreds of milliseconds, which is comparable to switching times of commercial black and white electrophoretic displays. Also, the device's durability was excellent. Within thousands of switching cycles, an intensity change less than 10% was observed without any precaution. However, the plasmonic colors predefined during the fabrication process could only be switched between on and off states. To realize a wider gamut, which is indispensable for full-color displays, the authors fabricated color pixels composed of micrometer-sized red, green and blue sub-pixels. Each sub-pixel contained the MIN films and a polyporol layer. As demonstrated in Fig. 5B, the authors could produce yellow, purple and cyan by mixing the appropriate red, green and blue colors provided by the sub-pixels. The absolute reflectivity and the contrast were very similar to ink-spots printed with a laserjet printer on ordinary papers. The similar visual appearance compared to standard color printed images in combination with low power consumption, high contrast and reflectivity, wide gamut, long-term stability, large scale production also on flexible substrates makes this approach suitable for full color electronic papers. Higher switching rates may be realized by thinner polymer films in the future.

While this electrochromic switching approach shows an impressive performance, potential future mass production may be limited by costs and availability of gold. In a subsequent study, the authors overcame this limitation by implementing abundant metals only in the MIN film (*70*). The multilayer system comprised a metal Al layer and two $Al_2O_3$ films, preventing a semi-transparent Cu film with short-range ordered nanoholes prepared by colloidal lithography from oxidation (see Fig. 5C). As in the initial work, a high reflection was achieved by interfering cavity modes of the $Al_2O_3$ spacer layer and the excitation of surface plasmons. Remarkably, a wide gamut was demonstrated by realizing static RGB pixel triplets in a plasmonic color print. To produce reflective colors dynamically, the MIN configuration was slightly modified. A 100 nm thick $SiO_2$ layer was deposited on top of the multilayer system to additionally protect copper from the electrolyte. As a switchable electrochromic polymer layer (500 nm thick), highly conductive PEDOT doped with polystyrenesulfonate (PSS) was screen-printed on the protection layer. The polymer film served as the functional medium and concurrently as an electrode due to its sufficient conductivity, also at the low-conductive state. An ITO covered plastic was used as a counter electrode, making the entire device flexible and bendable. By applying different voltages, the polymer could be switched from its conductive (strongly absorbing) state to its low-conductive (slightly absorbing state). Similar to the initial work, different plasmonic colors generated by the NIM could be switched on and off as well, however, with a moderate contrast only. The reason was twofold. First, the polymer's conductive state ("off") was not entirely dark but exhibited a blue tone. Second, PEDTOT:PSS was not fully transparent in the bright state ("on"), resulting in an reduced absolute reflectivity. Additionally, the switching times on the order of seconds were fairly large compared to the Au-based NIM devices. Also, the lifetime of the device defined by the maximum number of PEDOT switching cycles was lower. While such a performance may meet the criteria for simple reflective full-color electronic readers, there is still room for improvement. On the one hand, conductive polymers other than PEDOT provide a better switching contrast. On the other hand, the layer thickness and material of the protection layer, preventing Cu from oxidation, can be optimized. Thinner protection layers, for instance, will increase the interaction between the strongly confined enhanced plasmonic near-fields of the NIM structure and the electrochromic polymer. As a result, the thickness of the electrochromic polymer is reduced, leading to faster switching times as elaborated above. In summary, the advantage of mass production compatibility enabled by abundant materials comes along with a significant decrease in performance.

Even though the presented EC-based coloration concepts offer fast and high contrast switching between dark and distinct color states, they are not capable of true direct color tuning, *e.g.* from red to blue. In addition to the previously discussed sub-pixel approach, electrochromic polymers offer another route to dynamically generate such plasmonic colors tunable in the visible spectral range (*71*). PANI, for example, possesses three well-defined states (leucoemeraldine (fully reduced), emeraldine (half oxidized), and pernigraniline), with numerous intermediated oxidation states each featuring a different absorption. During voltage sweeping, the color of mere PANI could be continuously tuned from transparent yellow to green, to blue, to violet and *vice versa* as shown in Fig. 5D. Such a coloration capability has been recently combined with plasmonic nanostructures. Gold nanomeshs residing on AAO templates were fabricated and electrochemically coated with a 40 nm thick layer of PANI. Together with a reference electrode and a Pt electrode, the nanostructured electrode was immersed into an aqueous solution containing aniline and $HNO_3$. Cyclic voltammetry with a rate of 30 mV/s and reflection spectroscopy were performed simultaneously. During cycling, the vibrant colors were reversibly tuned with high contrast by the absorption variations of PANI. Remarkably, non-monotonous color changes were deduced from color images and reflectance spectra (Fig. 5D). At negative and low voltages (leucoemeraldine-like states), the color changed from brownish over yellowish to brownish again. Voltages higher than 0.2 V (emeraldine and pernigraniline-like states) led to a steep color switching from green to blue, violet, and black. This behavior substantially differed from PANI induced coloration of flat gold surfaces, where a monotonous color change from red over green to blue was observed. Since those changes occurred over a broad spectral range without any noticeable resonance-related features, the authors attributed the differences to a more general effect associated with the nanostructured environment. Possible explanations included a modified charge transfer kinetics introduced by the remarkably high surface areas of the nanostructures. This interface-related effect in combination with the highly confined plasmonic near-fields provided higher reflection contrast (70% between dark and color states) and a steeper color switching accompanied with faster switching times compared to PANI-coated flat gold films. Typical switching times for a full color switching cycle were on the order of several tens of seconds. In view of potential display devices, the switching times have to be drastically reduced, *e.g.* by optimization of the nanostructures.

So far, the potential of electrochromic polymers for dynamic plasmonic color generation has not been fully exploited yet. As a matter of fact, variations of the absorption coefficient are inherently accompanied by refractive index changes. Absorption changes of the EC polymers substantially modify the intensity of the plasmonic excitations and thereby alter the plasmonic colors. Simultaneously, refractive index variations of the EC polymers shift the resonance frequency of the plasmonic excitation and thus alter the associated colors as well. The perceived color is hence an indistinguishable mixture from both contributions. This fact becomes particularly important, if the refractive index and absorption are strongly dispersive as for alkoxy-substituted poly(3,4-propylenedioxythiophene) [PProDOT(CH$_2$OEtH$_x$)$_2$], termed ECP-M in the following, for example. Whereas a non-dispersive complex dielectric function could be desirable in view of targeted color control, dispersive materials potentially enrich the chromaticity. However, Ledin *et al.* demonstrated such refractive-index-induced LSPR shifts for gold nanorods (AuNRs) embedded in a functional EC polymer matrix as (see Fig. 5E) (*72*). The hybrid polymer-metal systems were fabricated by spray-coating an aqueous solution containing chemically synthesized gold nanorods onto an ITO substrate. Subsequently, a 40 nm thin electrically responsive ECP-M layer was spin-casted onto the nanorods. The specially designed electron-rich ECP-M possessed an excellent solubility in organic solvents and a low oxidation potential. More importantly, it could be electrically switched from a magenta color to a transparent state with high refractive index contrast as proven by ellipsometry. In addition to electrical switching, the plasmonic colors controlled by electrochromic polymers were also adjusted by the use of different pH values (*73, 74*). However, when Ledin *et al.* applied a voltage stepwise increased from −0.2 to 0.5 V to the ECP-M/AuNR system, the refractive index variation of ECP-M gradually blue-shifted the LSPR by 27 nm (see Fig. 5E). Simultaneously, the LSPR intensity was decreased by the emerging ECP-M polaron absorption during electrochemical oxidation. The modulation in intensity and resonance frequency was reversible as proven over 10 oxidation-reduction cycles. Additionally, a good qualitative agreement with FDTD simulations was found. The numerical calculations further showed that the blue shift strongly depended on the polymer's thickness, but it saturated for thicknesses exceeding the decay length of the highly confined plasmonic near-fields (roughly 11 nm). Accordingly, more complex plasmonic systems with higher sensitivities to refractive index changes, as for example shown by Zhang *et al.* (*75*), or more appropriate electrochromic polymers are required to extend the color tuning range. Recently, an LSPR shift of roughly 100 nm for PANI coated-gold nanorods have been reported, suggesting PANI as a promising candidate for refractive index-based dynamic plasmonic color generation (*76*). The core-shell configuration reported in this work and other works (*77–79*) is particularly attractive because of the optimized light matter interaction provided by a complete overlap of the enhanced electromagnetic fields surrounding the core nanoparticle and the electrochromic polymer shell.

Peng *et al.* advanced the concept and utilized the tunable refractive index of EC polymers to dynamically control plasmonic colors (*78*). In the experimental study, continuous color tuning was demonstrated on the single particle level (see Fig. 5F), providing the smallest-area active plasmonic pixels up to date, which could be potentially scaled up to wafer-sizes. Using surfactant-assisted chemical oxidative polymerization, the authors encapsulated colloidal gold nanoparticles (AuNPs) by a thin PANI shell and drop-cast them onto a planar Au substrate. The shell thickness defined the distance of the AuNP to the gold surface, resulting in highly confined electromagnetic hot spots and an additional coupled plasmonic mode (termed mode *c* in Fig. 5F). The coupled plasmonic resonance mode was highly sensitive to polarizability changes (refractive index change of $\Delta n = 0.6$) in the hot spot, *e.g.* introduced by the different PANI states. Immersed in an electrochemical cell, the Au-PANI core-shell nanoparticles on top of the Au mirror served as an electrode and the redox state of the PANI shell was adjusted by sweeping the voltage from −0.2 to 0.6 with a rate of 50m/s. In the fully reduced (leucoemeraldine state), the plasmonic band appeared at 642 nm in the simultaneously recorded dark-field scattering spectra. When the voltage was increased, the refractive index gradually changed and the LSPR blue-shifted by more than 100 nm, until PANI reached its fully oxidized (pernigraniline) state. The generated plasmonic colors ranging from red to green were vivid and could be reversibly switched with high contrast (50%) within 32 ms (oxidation) and 143 ms (reduction), respectively. Such a switching performance is already comparable to commercial video rates in current display devices. Moreover, the demonstrated device showed a high bistability (stable PANI charge state were retrained for more than 10 min), low energy consumption (even lower than that of commercial e-papers), and it was readily scalable. However, its gamut is considerably small. In view of potential full color display applications, this challenge has to be addressed, *e.g.* by the implementation of sub-pixels.

Electrochromic polymers and plasmonic nanostructures perfectly complement each other. Plasmonic nanostructures provide vibrant, but static colors with high resolution. Electrochromic polymers as functional materials offer a fast and readily refractive index switching enabled by electrically controlled redox reactions. Combined to hybrid systems, high contrast, excellent bistability, large gamut, high reflection efficiency, low power consumption, high resolution, long lifetime and mass production compatibility were demonstrated. Depending on the particular configuration, some of the performance parameters cannot compete with those of current technologies. Others are comparable or even exceed the performance of commercial display devices. Even though such impressive advances have been achieved, several challenges remain and require research efforts, *e.g.* a further improvement of the switching times, the addressability of a single-pixel or the simultaneous optimization of all performance parameters. To this end, the seemingly infinite variety of hybrid electrochromic polymer-plasmonic nanostructures has to be explored.

**Other functional materials**

Similar to electrochromic polymers, inorganic electrochromic transition metal oxides, such as tungsten trioxide ($WO_3$) or titanium oxide ($TiO_2$), dynamically change their optical properties through cyclic oxidation and reduction. During a redox-reaction, electrons and guest ions, such as H+ or Li+, are simultaneously injected into a redox-active host material, *e.g.* $WO_3$ (*66*). Consequently, the charge carrier distribution and thus the complex refractive index (including absorption and refractive index) are modulated to a large extent. Electrochromic transition metals provide numerous advantages over organic electrochromic materials, *e.g.* good thermal and chemical stability, long durability, solution-free operating, and last but not least good compatibility with standard microfabrication processes (*80*). Although electrochromic transition metals have found primary applications in thermal and light management for buildings and airplanes, they still suffer from low switching speeds, poor color-tuning versatility and low color efficiency. Implemented in plasmonic or cavity systems, these challenges could be addressed as recently demonstrated (*80*, *81*). Li *et al.*, for instance, fabricated metal-insulator-metal plasmonic resonators composed of a thin $Li_xWO_3$ spacer layer sandwiched between an Al layer and Al nanorods as shown in Fig. 6A (*80*). The gap plasmon and the associated plasmonic color, sensitively responded to changes of the $Li_xWO_3$ optical properties, which could be controlled by the Li concentration *x*. The all-solid device was operated at 80°C to increase the ionic conductivity. When a certain voltage was applied, Li ions were injected into the $Li_xWO_3$ layer from ionically connected $Li_yFePO_4$ ($y \sim 0.7$) electrodes nearby. Thereby, the refractive index of $Li_xWO_3$ could be switched from 2.1 in the lithiated state (V = −1.4 V) to 1.9 in the delithiated state (V = 1 V). As the refractive index changed, the resonance conditions were altered and a LSPR shift of 58 nm from ~620 nm (purple color) to 565 nm (blue color) was observed in the reflectance spectra. The benefit of the plasmonic structures was twofold. First, the plasmonic structures produced brilliant structural colors. Second, the enhanced light matter-interaction in the plasmonic hotspots allowed to substantially lower the $Li_xWO_3$ layer thickness down to 17 nm only. As a result, the parasitic absorption at non-resonant wavelengths was reduced to less than 5%. Additionally, the switching time, a crucial parameter in display devices, was improved to 20 s. The authors further demonstrated continuous color tuning upon cyclic voltammetry sweeping and a good bistability on the timescale of minutes. However, the approach is limited by its long switching times and high operation temperature of 80°C. Although an optimization of the doping process, *e.g.* the use of protons instead of Li$^+$ ions, may further improve the ion diffusivity and thereby the switching times, the presented configuration seems to be rather inapplicable to display technologies in the current stage.

In the presented Al/Li$_x$WO$_3$/Al layer system, the rather broadband plasmonic absorption was one of the limiting factors, which hampered subtle plasmonic color tuning. Fabry-Perot cavities possessing narrow resonances offer a solution to it. Recently, asymmetric FP nanocavities have been employed to demonstrate rich and subtle structural color tuning in reflection geometry (*82*). Such nanocavities were fabricated by successively sputtering uniform layers of tungsten and amorphous tungsten oxide on poly-ethylene terephthalate (PET) substrates. The fabrication process was rather simple, since it required no nanostructuring and was compatible with existing standard electrochromic fabrication processes. However, light incident onto the sample was reflected back and forth at the WO$_3$ interfaces, enhancing or suppressing the reflected light at specific wavelengths in dependence on the WO$_3$ layer thickness. As a result, reflectance modulations as large as 50% and various distinct structural colors could be produced, which remained almost invariant at oblique angles of incidence between 0° to 40°. By inserting Li ions provided by an external reservoir into the WO$_3$ layer, its refractive index was continuously altered, *e.g.* from 2.15 to 1.61 at a wavelength of 600 nm. Since the cavity resonance was directly linked to the WO$_3$ refractive index in this configuration, the FP resonance could be gradually shifted by more than 240 nm resulting in a wide color modulation from red (0.5 V) to green (−0.8 V) for a 163 nm thick WO$_3$ film. The shift was reversible, if Li was extracted from the WO$_3$ under the application of appropriate voltages. The reversible switching had a high coloration efficiency and exhibited a good cycling stability of more than 1000 cycles. Similar to other coloration approaches based on inorganic electrochromic materials, the switching time between a steady bleached and color state was on the order of a few seconds, which necessitates further improvement in view of display devices. The authors further improved the color-saturation by adding a metal layer on top of the active WO$_3$ layer (see Fig. 6B). As a demonstration, color butterfly images were prepared by photolithography on flexible substrates and switched between different color states. Obviously, a rich color tunability with a wide color gamut could be achieved.

Wu *et al.* introduced a CMOS-compatible technique to alter the absorption properties of functional dielectric $TiO_2$ metasurfaces for dynamic color generation (*30*). In contrast to the $WO_3$-based switching strategies, oxidation and reduction was realized by implanting $H^+$ ions and $O^-$ ions in an inductively coupled plasma etcher, respectively. Following this approach, the fabrication complexity could be reduced, since no additional ion reservoirs were required. Even though not plasmonic, $TiO_2$ metasurfaces, and dielectric metasurfaces in general, hold great potential for structural color generation (*13*). Other than that, $TiO_2$ metasurfaces have also been applied to optical devices, operating in the visible, *e.g.* metalenses with efficiencies above 70%, metasurface-based holograms, chromatism-corrected lenses, just to name a few (*83*). The optical properties of $TiO_2$, are ideally suited for such optical devices and moreover for reversible color generation. Its refractive index is high enough to support Mie resonances and its absorption can be continuously adjusted from transparent to black by controlling the ion implanting time. In contrast to $WO_3$-based coloration, where the plasmonic color was dynamically adjusted by refractive index variations, the $TiO_2$-based approach exploited the intrinsic absorption coefficient only. This is similar to the previously discussed Mg-based dynamic coloration, where the intrinsic optical properties of metallic Mg were modulated upon hydrogenation (*14*, *27–29*). However, using electron-beam-lithography and evaporation techniques, the authors fabricated a $TiO_2$ metasurface composed of periodically arranged $TiO_2$ nanoblocks (*30*). The coupling between the Mie resonances and the reflection of the periodic lattice gave rise to strong reflectance (up to 70%) and narrow full width at half maximum (FWHM) of 20 nm only (see Fig. 6C). As a result, vibrant colors covering the entire spectral range were generated by adjusting the periodicity and particle size in the fabrication process. After 4 minutes of implantation, $TiO_2$ metasurfaces were converted to absorbing (black) $TiO_2$ and the peak reflectance was reduced to less than 10%, resulting in a brownish color. The initial colors could be restored by 5 minutes of ion implementation with a good reproducibility, even after 20 cycles. The rather slow switching time requires significant improvements to become competitive with current and future display technologies. Additionally, the cumbersome plasma etcher instrumentation needed for switching hampers its applicability for displays.

Another metasurface-based strategy for dynamic structural color generation with even nanosecond switching times was demonstrated utilizing perovskites nano gratings as functional materials (*84*). In the past, perovskites have attracted great interest due to their dielectric properties and the multitude of applications in photovoltaics (*85*). The direct band-gap semiconductor methylammonium lead halide perovskite (MAPbX$_3$, with MA = CH$_3$NH$_3^+$ and X = Cl$^-$, Br$^-$, I$^-$ or mixtures), for example, possesses excellent photoluminescence (PL) properties including a high quantum efficiency and a narrow FWHM. It can emit green PL or even laser light with high intensities (intrinsic color) precisely adjustable by the optical excitation with a pump laser. Other colors, *e.g.* red or blue, can be generated by adjusting the stoichiometry of MAPbX$_3$ or via anion exchange. Additionally, MAPbX$_3$ has refractive indices between 2.1 and 2.5, which are high enough to support Mie resonances in a single perovskite nanostructure. Arranged in metasurfaces comprised of arrays with subwavelength periodicity, such perovskite nanostructures give rise to strong reflection (structural color) as recently demonstrated. Using electron beam lithography, the authors fabricated gratings of thin perovskites films as schematically illustrated in Fig. 6D. The distinct periodicity *P* of the array and the gaps between the MAPbX$_3$ stripes determined the structural color, which could easily be altered by changing the geometric parameters during the fabrication process. Illuminated by white light only, distinct structural colors predefined by the nano grating geometry, were generated, *e.g.* red for *P* = 382 nm and *d* = 110 nm. Remarkably, the intrinsic optical losses caused by the tailored band-gap were negligible and only scattering losses reduced the high reflection colors (40-65%). The intrinsic emission color (green) originating from the perovskite's band-gap was generated upon illumination with an Ti:sapphire laser. According to the color mixing theorem, which states that the hue can easily be tuned by varying the ratio of two colors, various colors were produced by mixing the intrinsic and structural colors. At low laser intensities, the intrinsic emission at 515 nm was minor, and the mixed color was dominated by the reflected structural color (red, 625 nm). For higher laser pumping densities, the green color became brighter and eventually dominated the spectrum. Consequently, the ratio of intrinsic (green) and structural (red) color was modulated, leading to different orange tones. The reversible switching offered ultrafast switching times on the order of nanoseconds defined by the photoluminescence lifetime. Such ultra-fast switching times exceed those of standard commercially available displays by orders of magnitudes. Additionally, spatially modulated laser beams may allow for arbitrary actuation of nanostructures and thereby provide an enhanced color control. However, more research is required to realize a wider gamut, *e.g.* by the use of other gain materials (GaN, ZnO *etc.*) or different stoichiometries of MAPbX$_3$. Conceptual shortcomings such as the laser's inherent high energy consumption and the limited compactness are further challenges to be addressed in future research. Despite non-plasmonic, both dielectric metasurface-based approaches are yet other examples for dynamic structural color generation, which may foster novel hybrid dielectric-metal concepts, when combined with plasmonic elements.

Thermochromatic materials offer another pathway to dynamically control plasmonic colors. $VO_2$, for example, transits from a monoclinic to a rutile phase at a critical temperature of 68 °C (*86*). The insulator to metal phase transition is accompanied by a substantial change of the electronic and optical properties. By switching the temperature from 20 °C to 80 °C, a thin $VO_2$ layer alters its color from green to yellow. Integrated in plasmonic nanostructures, additional colors were realized (*87*). Using electron beam lithography and evaporation techniques, a metal-insulator-metal (MIM) structure comprised of a $VO_2$ layer beneath a $SiO_2$ spacer layer and periodically arranged silver discs was fabricated (see Fig. 6E). During interaction with light, LSPRs, SPP and Wood's anomaly were excited. Depending on the periodicity, particle size and spacer layer thickness, distinct reflective colors could be generated across the entire visible spectral range. When the temperature was switched from 20 °C to 80 °C across the critical temperature, the $VO_2$ transformed from insulator to metal and the real part of the permittivity decreased, while the imaginary part increased. As a result, the reflection peak shifted to shorter wavelengths and the color was switched, *e.g.* from green to yellow. As shown in Fig. 6E, a color image comprised of five different MIM configurations with different distances and sizes was switched between two colors. Noteworthy, the perceived color was a mixture of the plasmonic color and the intrinsic $VO_2$ color. In addition to such mixed dull colors, long switching times on the order of hours and high operation temperatures hamper practicable applications. Even though the critical temperature can be potentially reduced by the exposure to hydrogen or introducing an electrical current, the method is still energy inefficient and complex for implementations in practicable display devices.

Liu *et al.* introduced a hydrolysis-based strategy to dynamically adjust the spacing between strongly interacting plasmonic nanostructures and thereby the plasmonic colors (*43*). In the experiments, a suspension of poly(acrylic acid)-coated (PAA-coated) silver nanoparticles was spray-coated on top of a layer of boron nitrate ($Na_2B_4O_7$) residing on a glass substrate. In the presence of water, $Na_2B_4O_7$ rapidly hydrolyzed into $H_3BO_3$ and $OH^-$ ions were released. The released ions simultaneously deprotonated the carboxyl groups of PAA attached to the silver nanoparticles. The deprotonation process increased the surface charges and concurrently the electrostatic repulsion between adjacent particles. As a result of this process, the distance between nanoparticles was increased and their plasmonic coupling weakened. A typical film (200 nm thick) containing silver nanoparticles with an average diameter of 8.5 nm exhibited a plasmonic band at 526 nm and an associated pink color (see Fig. 6F). If the sample was exposed to 80% relative humidity, the LSPR blue-shifted to 423 nm in 220 ms. During this process, the color was transformed from pink to red, red to orange and eventually orange to yellow. Remarkably, only slight distance modifications resulted in sustainably large blue-shifts on the order of 100 nm, demonstrating the benefit of coupled plasmonic nanostructures for dynamic plasmonic color control. When the moisture was completely removed, the coloration was reversed and the initial state could be reached in 640 ms. This relatively fast color switching was enabled by the rapid hydrolysis reaction and the thin layer thickness, which provided short diffusion lengths of $OH^-/H^+$ for the deprotonation and protonation of the PAA molecules. The hydrolysis-based approach is interesting due to its reversibility, ease of scalable fabrication and fast switching times. However, the humidity limits its application to display devices operating in real-word atmospheres with varying humidity. Here more research efforts, *e.g.* on device encapsulation or external stimuli other than moisture are required.

In summary, a large variety of functional materials and nanostructure designs have been implemented for dynamic structural color generation. Particularly, the combination of plasmonic coloration with other structural coloration strategies, such as dielectric metasurfaces or FP cavities, may lead to novel dynamic concepts. However, the majority of the conceived strategies are still in their infancy. More research endeavors are needed to optimize the current designs, characterize the performance and explore the potentials for dynamic structural color generation.

## 4. Discussion and conclusion

The rapidly growing research field of dynamic plasmonic color generation has attracted intense attention in the last several years. Starting from relatively straightforward demonstrations by electrochemically controlled size modulations of metallic nanostructures, the field has been dynamically developed and advanced. More and more sophisticated coloration concepts have been explored. A fast and efficient control of the vivid plasmonic colors enabled by functional materials has been successfully realized, even on a single pixel and nanostructure level. Moreover, first integrations into conventional transmissive LCD panels were reported (*54*), demonstrating the compatibility with current display technologies. From a nearly infinite number of functional materials, magnesium-hydrides, liquid crystals, and electrochromic polymers offered great potentials to dynamically control plasmonic colors and have been intensively studied. Other functional materials, such as perovskites or thermochromic vanadium dioxide, have been proposed for dynamic coloration as well. Some of these approaches showed excellent individual key performance indicators, *e.g.* a switching time on the order of nanoseconds or a high bistability, but they are lack of an acceptable overall performance. However, it is too early to evaluate the potential of these freshly emerging concepts for full color displays with real-world applications in the current stage.

The performance of dynamic plasmonic coloration based on electrochemical deposition and other functional materials, including magnesium-hydrides, liquid crystals, and electrochromic polymers is determined by the constituent plasmonic nanostructures as well as the functional materials. While the dynamic coloration control strongly depends on the integrated functional material, the excellent nanoscale resolution and the basically everlasting colors provided by the plasmonic nanostructures are essentially similar for all coloration concepts. The nanoscale resolution is mainly set by the type of plasmonic color generation, *e.g.* by LSPRs, SPPs, lattice excitations or others, and the technical implementation of pixels, *e.g.* as monopixels or sub-pixels. Compared to monopixel designs, where a single pixel potentially provides all colors, merged pixels composed of several sub-pixels, typically RGB triplets, reduce the resolution. Given the minimum pixels size of 10 µm × 10 µm or even below, the merged pixels are still on the order of 100 times smaller than those of current high resolution displays. However, an ultra-high resolution of 10,000 dpi can still be reached with plasmonic display devices (*54*, *69*). In contrast to the excellent resolution, the dynamic control of plasmonic colors strongly depends on the coloration concept and functional material (see Fig. 1E-H). For example, dynamic plasmonic color generation based on electrically controlled metal deposition onto predefined nanostructures offers a reasonable brightness and gamut, but suffers from slow switching times on the order of seconds and a low durability due to wear-and-tear and oxidation. Additionally, a selective control of micro- or even nanometer-sized pixels seems to be out of reach, at least in the current designs. Magnesium-hydride-based approaches show a similar performance, but micro-and nanoscale pixels can be selectively and locally addressed, if electrically controlled nanoscale hydrogen sources are employed instead of macroscopic gas cells. The excellent performance of LC-based coloration control, *e.g.* the fast switching times and the long-time stability, is superior to that of other approaches and is the closest to the criteria required for commercial video displays. On the contrary, dynamic plasmonic color generation enabled by electrochromic polymers possesses slower switching speeds, but an excellent bistability, potentially resulting in an ultra-low power consumption of the display device. This also shows that none of the functional materials is intrinsically "best". The applicability and suitability are correlated to the required display applications.

The application potential of plasmonic colors for reflective-type displays even exceeds that for transmissive-type displays, where LCDs, OLEDs and emerging quantum LEDs (QLEDs) (*88*) are the standard. Commercially available black-and-white as well as full-color reflective displays suffer from slow response times (typically hundreds of milliseconds), low reflection efficiencies and/or relatively low contrasts. Here plasmonic nanostructures offer a promising alternative, when combined with LCs or electrochromic polymers. Compared to LCs, electrochromic polymers are particularly appealing, since no additionally polarizing elements are required. Additionally, due to the enhanced plasmonic near-fields, nanometer thin layers of electrochromic polymers can already allow for an efficient dynamical control of the plasmonic colors. This comes along with significant advantages, *e.g.* a compact fabrication of displays on even flexible substrates, an ultra-low power consumption and switching speeds on the order of tens of milliseconds or even below. However, more research and engineering efforts have to be exerted to put electrochromic and liquid crystal based dynamic plasmonic coloration into practicable applications. Most importantly, the challenge to achieve selective and local control of micro- and nanoscale plasmonic pixels has to be addressed. Depending on the particular implementation, addressable pixels require additional layers and electrical connections, which will increase the fabrication complexity as well as fabrication costs, and concurrently reduce the display efficiency. Compared to advanced nanofabrication methods, such as EBL or FIB utilized in the majority of the presented studies, fabrication techniques compatible with mass production may not easily fulfill the high-quality conditions of plasmonic color displays, especially in terms of the high-resolution requirements.

Functional materials combined with metal nanostructures offer an exciting route to generate dynamic plasmonic colors. Owing to the nanostructures' sub-wavelength sizes and the excellent dynamic control provided by the integrated functional materials, full-color displays with unpreceded resolutions and color durability feature a rigorous perspective for useful applications in holographic displays (*89–92*), highly secure information encryption

(*90*), and naked-eye color indicators of temperatures (*93*), strain (*41*, *42*), moisture (*43*), biological substances (*94*), gases (*27*) or others (*95*, *96*).

**Acknowledgments**

**Funding:** The authors acknowledge funding from the European Research Council (Dynamic Nano) grant.

**Author contributions:** All authors wrote the manuscript.

**Competing interests:** The authors declare no competing interest.

**Data and materials availability:** All data needed to evaluate the conclusions in the paper are present in the paper. Additional data related to this paper may be requested from the authors.

**Figures**

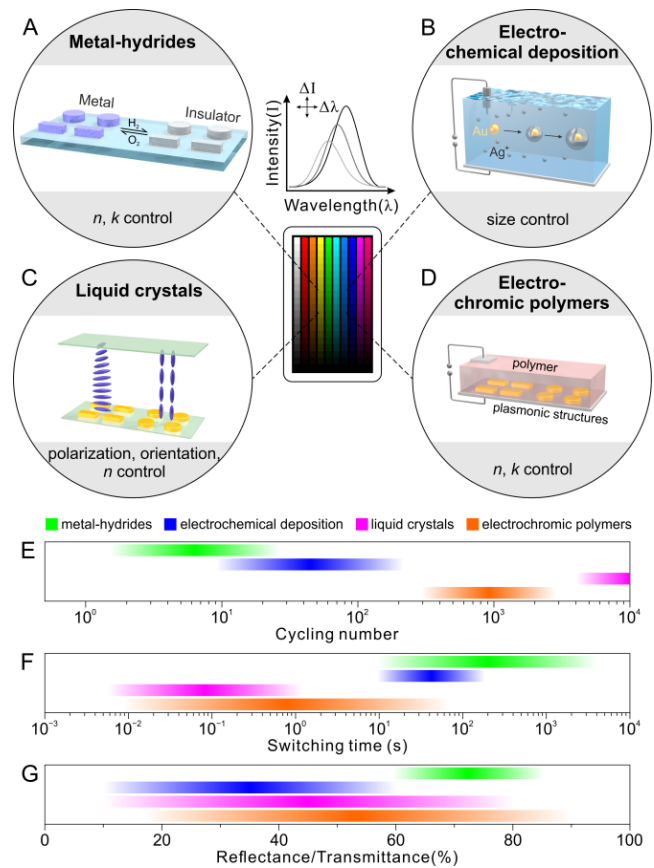

**Fig. 1. Dynamic plasmonic color generation enabled by functional materials and the related key performance indicators.** Among others, the intensity and resonance frequency of a plasmonic excitation determine the perceived plasmonic colors (middle panel). Both quantities can be effectively tuned through electrochemically induced size modulations of the constituent metal nanostructures, functional plasmonic materials themselves, or functional media surrounding the passive plasmonic elements. (**A**) Reversible transformations between metallic magnesium (Mg) and dielectric magnesium hydride (MgH$_2$) can take place upon hydrogen (H$_2$) and oxygen (O$_2$) exposures, respectively. The metal to insulator phase transition induces a change in the electron density, or more general, in the complex refractive index ($n + \text{i}\cdot k$) with $n$ and $k$ being the refractive index and the absorption coefficient, respectively. (**B**) Electrochemical deposition is applied to reversibly modulate the sizes of plasmonic nanoparticles. (**C**) Liquid crystals allow for a control of the polarization state of the incident or scattered light, the anisotropic refractive index $n$ of the liquid crystals, and the orientation of anisotropic nanoparticles embedded in the liquid crystals. (**D**) Switchable electrochromic materials surrounding the plasmonic nanostructures offer an efficient control of the complex refractive index.

Selected key performance indicators, such as (**E**) life time (cycling number), (**F**) switching time, and (**G**) reflectance/transmittance strongly depend on the coloration concepts.

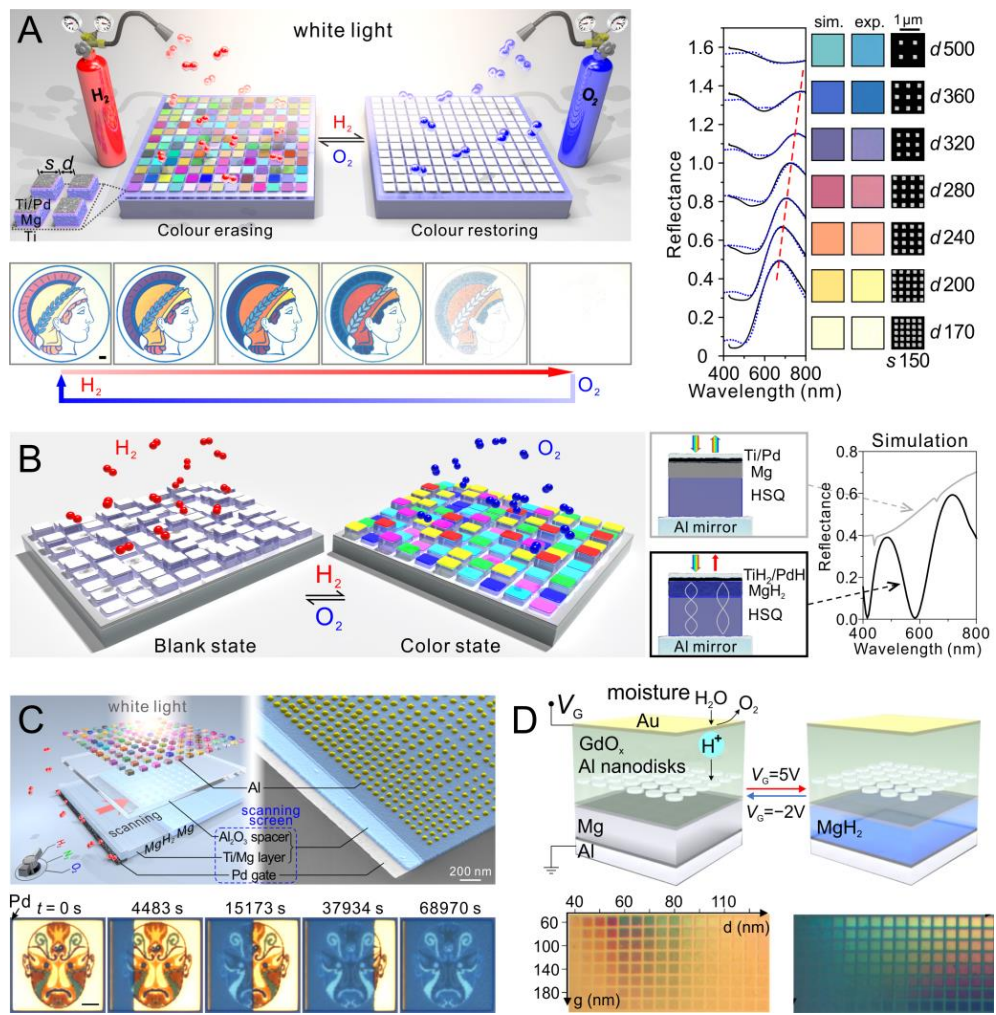

**Fig. 2. Hydrogenation of magnesium for dynamic color generation.** (**A**) Reflective plasmonic colors generated from Mg nanoparticles are erased (restored) upon hydrogen (oxygen) exposure as demonstrated for the Max-Planck Society's Minerva logo. Scale bar: 20 μm. Reflectance (spectra and colors) strongly depend on the nanoparticle size *s* and the interparticle distance *d*. (**B**) Structural colors emerging from Fabry-Perot cavities are reversibly switched between a color state and a blank state, when exposed to hydrogen and oxygen. (**C**) Schematic (left) and electron micrograph (right) of a scanning plasmonic color display. Upon hydrogen (or oxygen) loading, the Mg to $MgH_2$ (or $MgH_2$ to Mg) phase transition starts from the Pd gate and laterally evolves as shown for the Sichuan opera facial mask. Scale bar: 5 μm. (**D**) Illustration of an electrically controlled local proton source integrated in a plasmonic color generation device. When 5 V are applied, the injected hydrogen transforms Mg to MgH$_2$ resulting in color changes (see color palettes). (A) and (D) Adapted under the terms and conditions of the CC-BY Creative Commons Attribution 4.0 International License (*27*) and (*29*). Copyright 2017 and 2019, Macmillan Publishers Limited. (B) and (C) Adapted with permission from (*14*) and (*28*). Copyright 2017 and 2018 American Chemical Society.

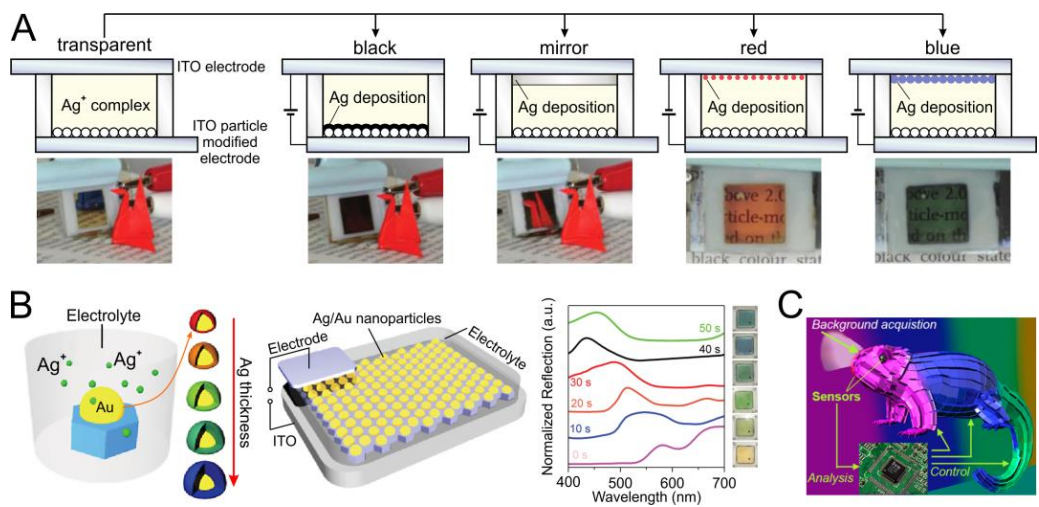

**Fig. 3. Electrochemical deposition of metals for dynamic color generation.** (**A**) Five different color states including transparent, black, mirror, red, and blue are realized by electrically controlled deposition of Ag on flat indium tin oxide (ITO) and ITO particle-modified electrodes. (**B**) Electrochemical controlled growth of Au-Ag core-shell nanoparticles in hexagonally arranged nanopores. As the Ag shell thickness increases with the deposition time, the plasmonic resonance shifts towards shorter wavelengths resulting in different plasmonic colors. (**C**) Demonstration of real-time light modulation for active camouflage: a mechanical plasmonic chameleon. Scale-like plasmonic color patches composed of $cm^2$-sized electrochemical cells mimic the background color. After sensing and spectral analysis, plasmonic colors are dynamically adjusted to the respective background colors using a reversible electrodeposition approach similar to **B**). (A) Adapted with permission from (*35*). Copyright 2013 WILEY-VCH Verlag GmbH & Co. (B) Adapted with permission from (*36*). Copyright 2014 WILEY-VCH Verlag GmbH & Co. (C) Adapted with permission from (*37*). Copyright 2016 American Chemical Society, https://pubs.acs.org/doi/10.1021/acsnano.5b07472. Further permissions related to the material excerpted should be directed to ACS.

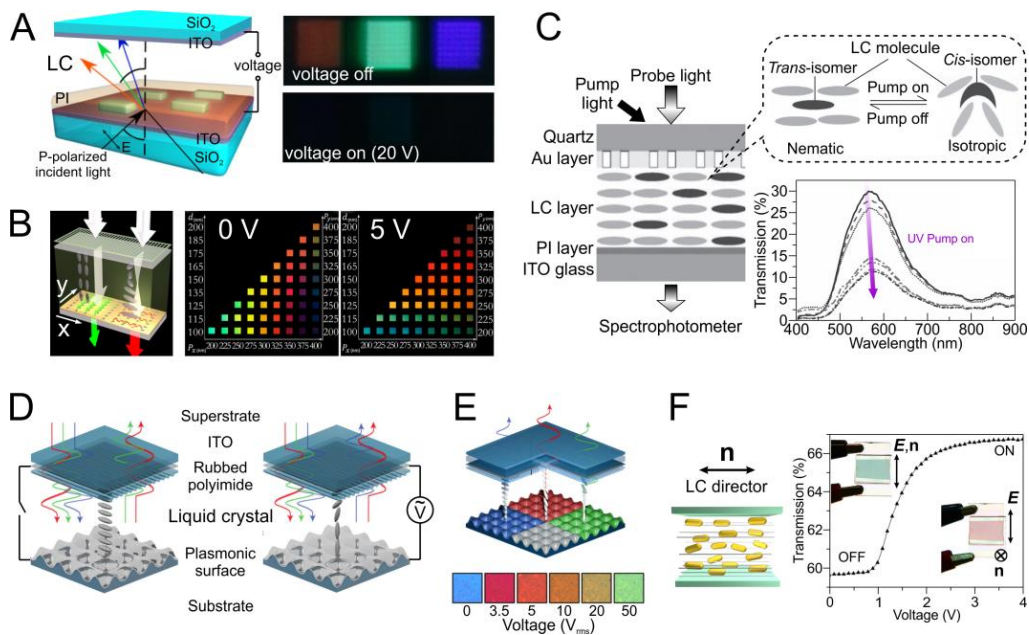

**Fig. 4. Liquid crystals (LCs) for dynamic color generation.** Plasmonic pixels (1.5 × 1.5 mm$^2$) composed of anisotropic nanorods covered by a LC layer are switched on and off by electrically rotating the polarization state of light. (**B**) Reversible switching between particular colors is achieved by modulating the polarization state of light incident on a LC-covered asymmetric-lattice metal nanohole array. (**C**) Optical switching of the anisotropic refractive index of LCs. (**D**) LC cell for electrically controlled refractive index modulation. (**E**) Electrically controlled polarization rotation and refractive index modulation of the LCs for plasmonic color generation. (**F**) Anisotropic gold nanorods embedded in a LC host are aligned by an external field ***E***. (A) and (B) Adapted with permission from (*47*) and (*46*). Copyright 2016 and 2017 American Chemical Society. (C) Adapted with permission from (*56*). Copyright 2012 WILEY-VCH Verlag GmbH & Co. (D) and (E) Adapted under the terms and conditions of the CC-BY Creative Commons Attribution 4.0 International License (*58*) and (*54*). Copyright 2015 and 2017, Macmillan Publishers Limited. (F) Adapted with permission from (*62*). Copyright 2014 American Chemical Society.

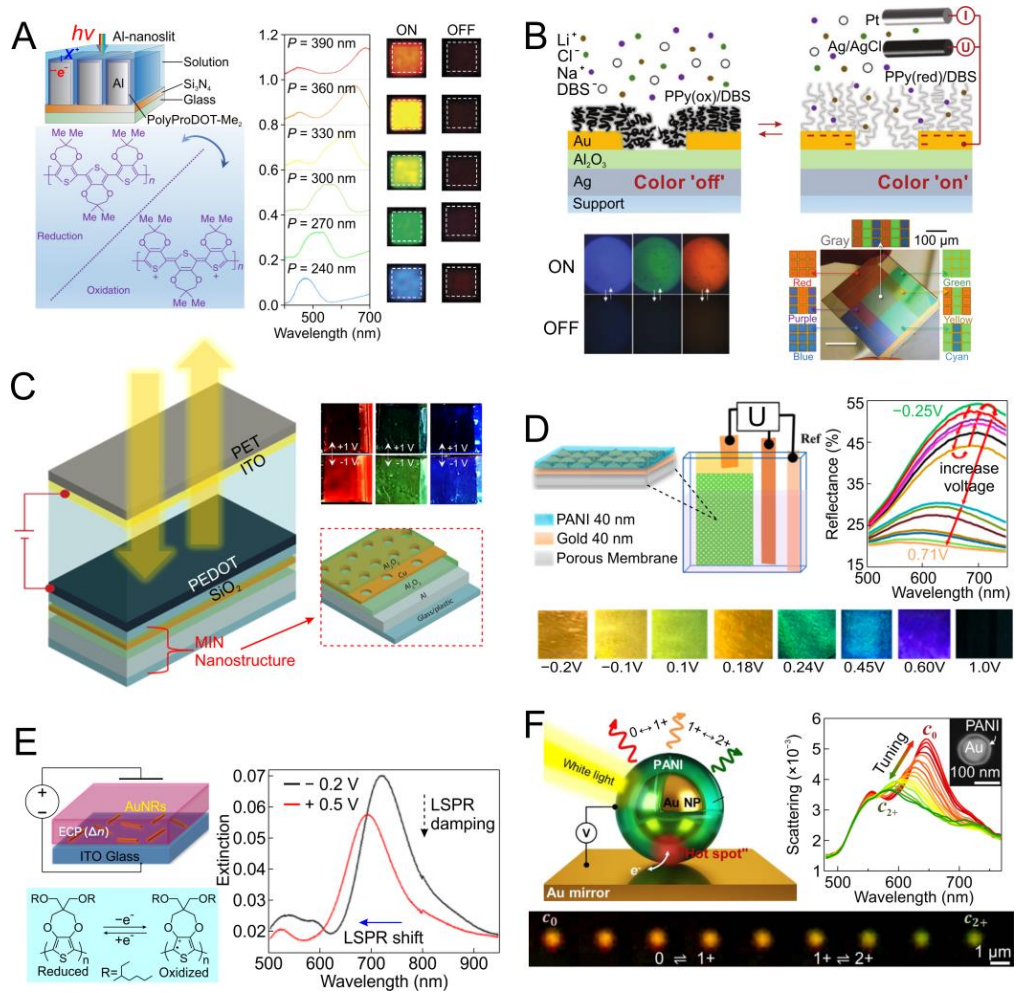

**Fig. 5. Electrochromic polymers (ECPs) for dynamic color generation.** (**A**) Electrochromic PolyProDoT-Me$_2$ coated on Al nanoslit arrays is electrically switched between a reduced (ON) state and an oxidized (OFF) state. (**B**) Depending on the redox state of polypyrrole (PPy) plasmonic colors are either absorbed by (OFF) or transmitted through (ON) the electrochromic polymer. (**C**) Cell design for reflective plasmonic color control using CMOS-compatible materials only. (**D**) Continuous, but non-monotonous plasmonic color tuning of a polyaniline-coated (PANI-coated) gold nanomesh. (**E**) During electrically controlled redox-reactions of an ECP, the LSPR is modulated in resonance frequency and in intensity. (**F**) PANI covered single gold nanoparticle residing on a gold mirror for dynamic plasmonic color generation. (A) Adapted under the terms and conditions of the CC-BY Creative Commons Attribution 4.0 International License (*68*). Copyright 2016, Macmillan Publishers Limited. (B) Adapted with permission from (*69*). Copyright 2016 WILEY-VCH Verlag GmbH & Co. (C),(D) and (E) Adapted with permission from (*70*), (*71*) and (*72*). Copyright 2017, 2019 and 2016 American Chemical

Society. (F) Adapted from (*78*). © The Authors, some rights reserved; exclusive licensee American Association for the Advancement of Science. Distributed under a Creative Commons Attribution NonCommercial License 4.0 (CC BY-NC) http://creativecommons.org/licenses/by-nc/4.0/.

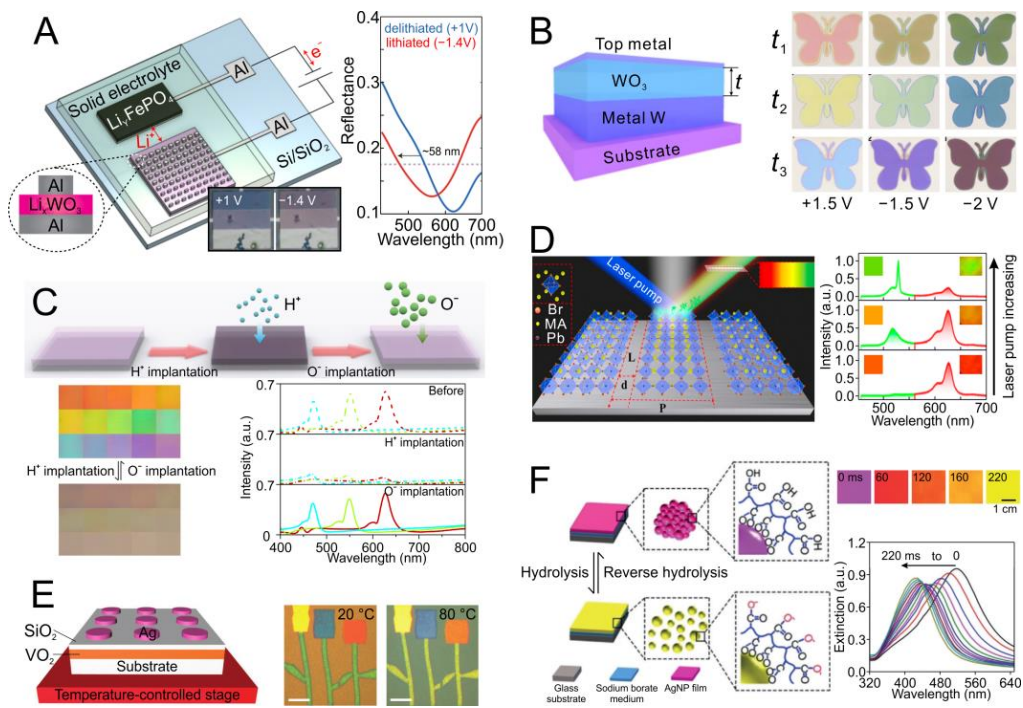

**Fig. 6. Other functional materials for dynamic color generation.** (**A**) A solid-state electrochromic device with Li$_x$WO$_3$ as functional material. (**B**) Fabry-Perot-type nanocavity filled with the functional material WO$_3$. Right panels: three butterfly-like electrochromic devices with sizes of several cm$^2$. (**C**) Structural colors generated by dielectric TiO$_2$ metasurfaces are reversibly switched. (**D**) Various colors are generated by mixing the intrinsic photoemission of perovskites and the structural color provided by the perovskite nanograting. (**E**) Printed flowers composed of thermochromic VO$_2$ films and metal nanostructures. Scale bar: 40 µm. (**F**) (Reverse) hydrolysis is employed to modulate nanometer-sized interparticle spacing. (A) Adapted with permission from (*80*). Copyright 2019 American Chemical Society. (B) Adapted under the terms and conditions of the CC-BY Creative Commons Attribution 4.0 International License (*82*). Copyright 2020, Macmillan Publishers Limited. (C) Adapted from (*30*). © The Authors, some rights reserved; exclusive licensee American Association for the Advancement of Science. Distributed under a Creative Commons Attribution NonCommercial License 4.0 (CC BY-NC) http://creativecommons.org/licenses/by-nc/4.0/. (D) Adapted with permission from (*84*). Copyright 2018 American Chemical Society, https://pubs.acs.org/doi/10.1021/acsnano.8b02425. Further permissions related to the material excerpted should be directed to ACS. (E) and (F) Adapted with permission from (*87*) and (*43*). Copyright 2018 and 2019 WILEY-VCH Verlag GmbH & Co.